\documentclass{jpp}
\usepackage{graphicx}

\usepackage[utf8]{inputenc}
\usepackage[T1]{fontenc}
\usepackage{amsmath}

\usepackage{pgf}

\usepackage{tikz}
\usepackage{pgfplots}
\usepgfplotslibrary{groupplots}
\usetikzlibrary{positioning}
\pgfplotsset{compat=1.18}

\usepackage[colorlinks=true, linkcolor=blue, citecolor=blue]{hyperref}
\makeatletter
\NewDocumentCommand{\blackauthorscite}{m s o o m}{%
	\begingroup
	\let\originalnameformat\NAT@nmfmt
	\renewcommand{\NAT@nmfmt}[1]{%
		\textcolor{black}{\originalnameformat{##1}}}%
	\IfBooleanTF{#2}
	{\def\citationcommand{#1*}}
	{\def\citationcommand{#1}}%
	\IfNoValueTF{#3}
	{\citationcommand{#5}}
	{\IfNoValueTF{#4}
		{\citationcommand[#3]{#5}}
		{\citationcommand[#3][#4]{#5}}}%
	\endgroup
}
\makeatother

\NewDocumentCommand{\mycitet}{}{\blackauthorscite\citet}
\NewDocumentCommand{\mycitep}{}{\blackauthorscite\citep}
\usepackage[nameinlink,noabbrev]{cleveref} 
\crefname{appendix}{Appendix}{Appendices}
\crefformat{equation}{(#2#1#3)} 
\crefrangeformat{equation}{#3(#1)#4—#5(#2)#6}
\crefformat{section}{#2#1#3} 

\shorttitle{Generation of low-frequency flows by Alfvén waves in MHD}
\shortauthor{J. M. Edmiston and others}

\title{Generation of low-frequency flows by shear Alfvén waves in magnetohydrodynamics}

\author{James M. Edmiston\aff{1,}\aff{2,}\aff{3}
  \corresp{\email{james.edmiston@physics.ox.ac.uk}},
  M. Fitzgerald\aff{2},
  J. Ruiz Ruiz\aff{1,}\aff{4},
  D. A. Uzdensky\aff{1},
 \and A. A. Schekochihin\aff{1,}\aff{5}}

\affiliation{\aff{1}Rudolf Peierls Centre for Theoretical Physics, University of Oxford, Oxford OX1 3PU, UK
\aff{2}UKAEA Fusion Association, Culham Science Centre, Abingdon OX14 3DB, UK
\aff{3}Magdalen College, Oxford OX1 4AU, UK
\aff{4}York Plasma Institute, University of York, Heslington, York YO10 5DD, UK
\aff{5}Merton College, Oxford OX1 4JD, UK}

\begin{document}

\maketitle

\begin{abstract}
Coherently forced, weak reduced-magnetohydrodynamic turbulence is studied in a uniform plasma to investigate the long-term nonlinear interactions of shear Alfvén waves. The `pump waves' are chosen to model toroidal Alfvén eigenmodes, excited by suprathermal particles in magnetic fusion devices. Modes with zero frequency and no variation along the magnetic field are generated nonlinearly; these produce analogues of zonal flows and currents in a gradient-free system. This indicates that quasistationary flows can be generated by shear Alfvén waves in systems simpler than those previously considered, which relied on compressibility, finite-Larmor-radius effects, or complex geometry. The saturated amplitudes of the modes are found analytically and numerically as a function of pump-wave amplitude. This facilitates estimates of the critical amplitude above which toroidal Alfvén eigenmodes can generate sufficiently strong zonal flows to suppress microturbulence in the core of magnetic fusion devices. This threshold corresponds to magnetic fluctuations whose amplitude, relative to the background field, is comparable to $\rho_{\star i}$, the ion Larmor radius normalised to the minor radius of the device.
\end{abstract}

\section{Introduction}

In the core of magnetic fusion devices, transport due to microturbulence determines the achievable performance. Intensive efforts over several decades have generated a detailed understanding of turbulent transport in non-burning plasmas \mycitep{horton_drift_1999, tynan_review_2009}. However, at higher densities and temperatures, the highly energetic alpha particles generated by fusion reactions can alter the confinement properties of the core. Such `burning plasmas' are approaching experimental reality as  SPARC \mycitep{creely_overview_2020}, BEST \mycitep{Yan2025BEST}, and ITER \mycitep{holtkamp_overview_2007} undergo construction, and even more ambitious experiments are planned for the future \mycitep{Creely_2026,chapman2024spherical,Anderson_Canik_Hegna_Mowry_2025}. Understanding the novelties of the burning-plasma state is therefore essential for accurate predictions of future reactor performance.

In addition to providing self-heating, the alpha particles directly impact the microturbulence via a number of mechanisms \mycitep[see reviews by][]{citrin_overview_2023,na_how_2025,mazzi_fast_2026}. Many of these effects are linear and expected to be small in future devices \mycitep{wilkie_first_2018}. However, transport may be appreciably suppressed by toroidal Alfvén eigenmodes (TAEs) \mycitep{cheng_high-n_1985,cheng_lown_1986}, which are shear Alfvén waves (SAWs) \mycitep{alfven1942existence} with discrete frequencies excited by fast ions in toroidal geometry. These modes have been observed alongside improved plasma performance both experimentally and numerically \mycitep{di_siena_electromagnetic_2019, mazzi_enhanced_2022}, ostensibly because they mitigated curvature-driven ion-temperature-gradient (ITG) turbulence in the core \mycitep{waltz_threedimensional_1988}.

The large frequency separation between the `fast' Alfvénic physics and `slow' ITG dynamics precludes any significant linear coupling. To influence the transport, TAEs are believed to produce zero-frequency `zonal modes' via a nonlinear mechanism \mycitep{di_siena_electromagnetic_2019}. Zonal modes are the quasistationary components of a field that are constant within a flux surface, but vary in the radial direction. Such modes in the electrostatic potential have radial electric fields that create $E \times B$ flows (`zonal flows') in the binormal direction. These then shear apart ITG eddies, shorten their radial correlation length, and hence reduce transport \mycitep{hammett_developments_1993}. Their posited generation by TAEs is supported experimentally by observations of zero-frequency modes phase-locked to the TAEs \mycitep{ruiz_ruiz_measurement_2025}, and transport-barrier formation due to flow shear at the TAE radial location \mycitep{du_microturbulence_2025}. In contrast, high-frequency flows (e.g., due to the TAEs themselves) would not give significant suppression, as their shearing effect would average to zero over the relevant ITG timescales \mycitep{hahm_shearing_1999}.

The key physics is therefore the nonlinear generation of low-frequency modes by high-frequency pump waves. The situation is complicated by the nature of the waves involved: in uniform and incompressible magnetohydrodynamic (MHD) plasmas, only counterpropagating SAWs can interact nonlinearly \mycitep{walen1944theory,elsasser_hydromagnetic_1950}. As discussed later, this might appear to preclude the generation of low-frequency modes constant along the magnetic field. For this reason, theoretical analysis is usually undertaken within more complex plasma descriptions, where compressibility, finite-Larmor-radius effects, or toroidal geometry introduce a small imbalance between the Reynolds and Maxwell stresses of the pump wave \mycitep{chen_nonlinear_2013,chen_physics_2016,qiu_nonlinear_2017, chen_effects_2024}\footnote{Such work originally aimed to determine whether zonal modes can alter the TAE saturation amplitude via `wave-wave' nonlinearity, rather than their potential impact on microturbulence.}. Initial growth rates of the modes have been derived, but quantitative predictions of the turbulence-suppression effect require predictions of the saturated amplitudes of the zonal flows. Thus, the effect of plasma burning on the core performance of future reactors remains uncertain.

In this work, we aim to elucidate the most fundamental mechanism of zonal-flow generation by SAWs, which can take place in a uniform, incompressible MHD plasma. TAEs have well-defined frequencies and small amplitudes ($\delta B_\perp/B \sim 10^{-3}$), so in isolation would coherently force weak MHD turbulence. While of significant theoretical interest \mycitep{ng_scaling_1997, goldreich_magnetohydrodynamic_1997, GALTIER_NAZARENKO_NEWELL_POUQUET_2000, perez_weak_2007, schekochihin_weak_2012, schekochihin_mhd_2022}, weak MHD turbulence is rarely considered to be relevant to real systems \mycitep[see, however,][]{saur2002evidence}, and unlike in the present paper, it is usually studied numerically with broadband forcing. In a reduced-MHD plasma \mycitep{kadomtsev_nonlinear_1973, strauss_nonlinear_1976}, we find that `zonal' modes with $k_\parallel = 0$ and $\omega = 0$ are generated on nonlinear timescales. These are the closest analogues to zonal modes in a plasma without equilibrium gradients, where the binormal and radial directions are equivalent. The shear-Alfvén pump waves in our system are generated by a simple antenna, which does not attempt to replicate TAE saturation via the `wave-particle' paradigm \mycitep{berk_saturation_1990}. Our emphasis is instead on studying the saturated modes produced by TAE-like pump waves of a given amplitude, via the MHD `wave-wave' nonlinearity. The resulting saturation levels are predicted analytically and confirmed by simulation. This enables an analytical \textit{a priori} estimate of the critical TAE amplitude for core transport suppression. Future work will aim to move towards a more realistic plasma model by reintroducing gradients and drift-wave turbulence, with the emphasis remaining on the saturated state and on obtaining quantitative predictions of the turbulence-suppression effect.

In \S\cref{sec:antennarMHD}, we describe a simple system within which a TAE-like antenna forces pump waves in reduced MHD, before then deriving a reduced model in \S\cref{sec:analmethods} that contains the essential dynamics of the system. This allows us to derive and discuss scalings for the saturated amplitudes in \S\cref{sec:analresults}, which are verified numerically in \S\cref{sec:numver}. We assess the implications for magnetic fusion devices in \S\cref{sec:discussion}, and summarise our conclusions in \S\cref{sec:conclusion}.

\section{Reduced MHD with a TAE-like antenna} \label{sec:antennarMHD}

In this section we will introduce our representation of the system as forced reduced MHD. First, we will present the equations of reduced MHD in \S\cref{sec:rmhd}, then motivate and derive a TAE-like Alfvén antenna in \S\cref{sec:antmotiv} and \S\cref{sec:antderiv}. This will be followed by a short discussion in \S\cref{sec:freqdoubl} of why modes at $\omega = 0$ and $k_\parallel = 0$ were previously believed to be absent in such a system. The essential dynamics will then be extracted in \S\cref{sec:analmethods} to produce a reduced model suitable for further analysis.

\subsection{Forced reduced MHD} \label{sec:rmhd}

We consider a simple homogeneous plasma in a 3D rectangular slab, with periodic boundary conditions. Employing $(x,y,z)$ coordinates, the system is characterised by a strong background magnetic field $\mathbf{B}_0 = B_0\hat{\mathbf{z}}$ that is uniform, static, and straight. To describe the plasma, we use reduced MHD (RMHD) to isolate SAW dynamics. This is a subsidiary limit of MHD in which parallel wavenumbers $k_\parallel$ along the magnetic field are ordered much smaller than perpendicular wavenumbers $k_\perp$, and fluctuations are ordered smaller than the background by the same amount. These ordering assumptions produce an incompressible system, where the perpendicular fluctuations in the velocity $\delta \mathbf{u}_\perp$ and magnetic field $\delta \mathbf{B}_\perp$ are decoupled from their parallel counterparts. These perpendicular fields are solenoidal ($\nabla_\perp \cdot \delta \mathbf{u}_\perp = \nabla_\perp \cdot \delta \mathbf{B}_\perp = 0$), allowing a reformulation in terms of scalar stream functions:

\begin{equation}
\delta \mathbf{u}_\perp = \hat{\mathbf{z}} \times \nabla_\perp \Phi, \quad v_\mathrm{A} \frac{\delta \mathbf{B}_\perp}{B_0} = \hat{\mathbf{z}} \times \nabla_\perp \Psi,
\end{equation} where we have used Gaussian units to define the Alfvén speed $v_\mathrm{A} = B_0/\sqrt{4 \pi \rho_0}$, with~$\rho_0$ the equilibrium mass density. We also note that the stream functions $\Phi$ and $\Psi$ are respectively proportional to the electrostatic potential $\phi = \Phi B_0/c$ and the parallel vector potential $A_\parallel = - B_0 \Psi / v_\mathrm{A}$. Including also forcing in both the magnetic field ($F_\Psi$) and the velocity field ($F_\Phi$), the governing RMHD equations become \mycitep{kadomtsev_nonlinear_1973, strauss_nonlinear_1976}
\begin{align} 
    \partial_t \Psi + \{\Phi, \Psi\} &= v_\mathrm{A} \partial_z \Phi + F_\Psi, \label{eq:Psieq} \\
   \partial_t \nabla_\perp^2 \Phi + \{\Phi, \nabla_\perp^2 \Phi\} &= v_\mathrm{A} \partial_z \nabla_\perp^2 \Psi + \{\Psi, \nabla_\perp^2 \Psi\} + \nabla_\perp^{2} F_\Phi,  \label{eq:lapphi}
\end{align} where we have defined the Poisson bracket $\{f,g\} \equiv \hat{\mathbf{z}} \cdot (\nabla_\perp f \times \nabla_\perp g) = (\partial_x f) (\partial_y g) - (\partial_x g) (\partial_y f)$. These equations are valid at arbitrary $\beta = c_s^2/v_\mathrm{A}^2$ (where $c_s$ is the sound speed), provided that both the Mach number $\mathrm{Ma} = \delta u_\perp / c_s$ and the Alfvénic Mach number $\mathrm{Ma}_\mathrm{A} = \delta u_\perp / v_\mathrm{A}$ associated with the fluctuations remain small. The relative size of $c_s$ and $v_\mathrm{A}$ is only important in the pressure term of the momentum equation, which is eliminated by taking the curl to find~\cref{eq:lapphi}.

As $\{\Phi, f\} = \hat{\mathbf{z}} \cdot (\nabla_\perp \Phi \times \nabla_\perp f) = (\hat{\mathbf{z}} \times \nabla_\perp \Phi) \cdot \nabla_\perp f = \delta \mathbf{u}_\perp \cdot \nabla_\perp f$, the left-hand sides of \cref{eq:Psieq} and \cref{eq:lapphi} are convective derivatives of $\Psi$ and the vorticity $\nabla_\perp^2\Phi$, respectively. Similarly, $v_\mathrm{A} \partial_z \nabla_\perp^2 \Psi + \{\Psi, \nabla_\perp^2 \Psi\}$ describes wave propagation along the total magnetic field $B_0\hat{\mathbf{z}} + \delta \mathbf{B}_\perp$. Dissipation terms due to viscosity and resistivity are neglected in our analytical model, on the grounds that we are interested in the dynamics on the scales where linear propagation, forcing, and nonlinearity compete. More explicitly, we assume that the dissipation time $\tau_\nu$ is much longer than both the linear time $\tau_\mathrm{A} \sim (k_\parallel v_\mathrm{A})^{-1}$ and the nonlinear time $\tau_\mathrm{nl} \sim (k_\perp^2 \Phi)^{-1} \sim (k_\perp^2 \Psi)^{-1}$ at the scales of interest. In our numerical work, all dissipation will be perpendicular fourth-order hyperdissipation, such that $\tau_\nu \sim (\nu k_\perp^4)^{-1}$, where $\nu$ is a very small constant. To ensure that the system is weakly turbulent ($\tau_\mathrm{A} \ll \tau_\mathrm{nl}$), pump-wave amplitudes are set to be small, as, for a TAE, $k_\perp \Psi / v_\mathrm{A} \sim \delta B_\perp / B_0 \lesssim 10^{-3}$, while $k_\parallel/k_\perp \gg 10^{-3}$.  The relationship between $\tau_\mathrm{A}$ and $\tau_\mathrm{nl}$ will be discussed in greater detail in \S\cref{sec:analmethods}. 

Rewriting \crefrange{eq:Psieq}{eq:lapphi} in terms of the Elsasser potentials $\zeta^\pm \equiv \Phi \pm \Psi$, we find

\begin{equation} \label{eq:rmhdzeta}
    \frac{\partial \zeta^{\pm}}{\partial t} \mp v_\mathrm{A} \partial_z \zeta^\pm = N[\zeta^\mp, \zeta^\pm] + F^\pm.
\end{equation} Here $F^\pm \equiv F_\Phi \pm F_\Psi$ is the stream function of a body force acting within the plasma, and the nonlinearity takes the form

\begin{equation} \label{eq:nonlin}
    N[\zeta^\mp, \zeta^\pm] \equiv \nabla_\perp^{-2} \left( 
\{\partial_j \zeta^\pm, \partial_j \zeta^\mp\} - \{ \zeta^\mp, \nabla^2_\perp \zeta^\pm \} \right),
\end{equation} where $\nabla_\perp^{-2}$ is the inverse of the perpendicular Laplacian $\nabla_\perp^2 = \partial_x^2 + \partial_y^2$, and a sum over $j \in \{x, y \}$ is assumed. We note that the somewhat complicated form of the nonlinearity is not a challenge. Linearising \cref{eq:rmhdzeta} reveals that the dispersion relations of the two fields $\zeta^\pm$ are $\omega_\mathrm{l}^\pm = \mp k_\parallel v_\mathrm{A}$. Consequently, the Elsasser potentials propagate in opposite directions, and the nonlinearity \cref{eq:nonlin} acts purely between counterpropagating waves, as previously anticipated. Only two other properties are required in the ensuing analysis: $N[\zeta^\mp, \zeta^\pm]$ is bilinear, and contains only perpendicular derivatives.

\subsection{Motivation of forcing} \label{sec:antmotiv}

As a minimal proxy for toroidal Alfvén eigenmodes, we add an external antenna to drive SAWs continuously. In a real device, a TAE is composed of Fourier harmonics with a single toroidal mode number $n$ but two adjacent poloidal mode numbers $m$ and $m+1$. The eigenmode occurs in a radial region where the two Fourier components would have the same frequency, such that $\omega \approx k_{\parallel,m} v_\mathrm{A} \approx - k_{\parallel,m+1} v_\mathrm{A}$, with the parallel wavenumber defined as $k_\parallel \equiv (nq-m)/qR$, where $R$ is the major radius of the magnetic axis and $q$ is the safety factor, which
 must take the value $(m+1/2)/n$ in the overlap region. Comparing with the generic linear dispersion relation $\omega_\mathrm{l}^\pm = \mp k_\parallel v_\mathrm{A}$, we identify the components of the TAE as counterpropagating waves with different perpendicular wavevectors, and parallel wavenumbers of equal magnitude but opposite sign.

\begin{figure}
    \centering
    \definecolor{diagramAntiqueWhite}{RGB}{250,235,215}
    \resizebox{0.8\linewidth}{!}{%
        \begin{tikzpicture}[x=1cm,y=1cm]
            \path[use as bounding box] (-0.254,-0.254) rectangle (16.254,9.254);

            \fill[diagramAntiqueWhite] (4,0.9) rectangle (12,8.1);

            \foreach \y in {
                1.453846,2.007692,2.561538,3.115385,3.669231,4.223077,
                4.776923,5.330769,5.884615,6.438462,6.992308,7.546154
            } {
                \draw[red,line width=1pt] (4,\y) -- (12,\y);
                \fill[red] (8.075,\y) -- (7.925,\y+0.075) -- (7.925,\y-0.075) -- cycle;
            }

            \draw[black,line width=2pt] (4,0.9) rectangle (12,8.1);

            \node[red,anchor=south west,font=\fontsize{17.28}{20.736}\selectfont]
                at (7.76,8.172) {$\mathbf{B}_0$};

            \draw[black,line width=1.8pt,smooth,domain=0:1,samples=300,variable=\t]
                plot ({1+2.5*\t-0.057582*sin(1080*\t)}, {2.25+0.9*\t+0.159951*sin(1080*\t)});
            \fill[black] (3.829311,3.268552) -- (3.452579,3.281724) -- (3.547421,3.018276) -- cycle;

            \draw[black,line width=1.8pt,smooth,domain=0:1,samples=300,variable=\t]
                plot ({1+2.5*\t+0.057582*sin(1080*\t)}, {6.75-0.9*\t+0.159951*sin(1080*\t)});
            \fill[black] (3.829311,5.731448) -- (3.547421,5.981724) -- (3.452579,5.718276) -- cycle;

            \draw[black,line width=1.8pt,smooth,domain=0:1,samples=300,variable=\t]
                plot ({15-2.5*\t-0.057582*sin(1080*\t)}, {2.25+0.9*\t-0.159951*sin(1080*\t)});
            \fill[black] (12.170689,3.268552) -- (12.452579,3.018276) -- (12.547421,3.281724) -- cycle;

            \draw[black,line width=1.8pt,smooth,domain=0:1,samples=300,variable=\t]
                plot ({15-2.5*\t+0.057582*sin(1080*\t)}, {6.75-0.9*\t-0.159951*sin(1080*\t)});
            \fill[black] (12.170689,5.731448) -- (12.547421,5.718276) -- (12.452579,5.981724) -- cycle;

            \node[anchor=north,font=\fontsize{17.28}{20.736}\selectfont]
                at (2.425,2.200049) {$\mathbf{k}_{\perp 2}$};
            \node[anchor=north,font=\fontsize{17.28}{20.736}\selectfont]
                at (2.425,5.860049) {$\mathbf{k}_{\perp 1}$};
            \node[anchor=north,font=\fontsize{17.28}{20.736}\selectfont]
                at (13.575,2.260049) {$\mathbf{k}_{\perp 1}$};
            \node[anchor=north,font=\fontsize{17.28}{20.736}\selectfont]
                at (13.575,5.740049) {$\mathbf{k}_{\perp 2}$};

            \draw[black,line width=1pt] (7.4,0.3) -- (8.6,0.3);
            \fill[black] (8.6,0.3) -- (8.42,0.39) -- (8.42,0.21) -- cycle;
            \node[anchor=west,font=\fontsize{17.28}{20.736}\selectfont]
                at (8.75,0.3) {$\hat{\mathbf{z}}$};
        \end{tikzpicture}%
    }
    \caption{Schematic of the physical model under consideration. Counterpropagating shear Alfvén waves with $\mathbf{k}_{\perp1}$ and $\mathbf{k}_{\perp2}$ are excited by an antenna, and used to force turbulence coherently in a uniform plasma.}
    \label{fig:diagram}
\end{figure}
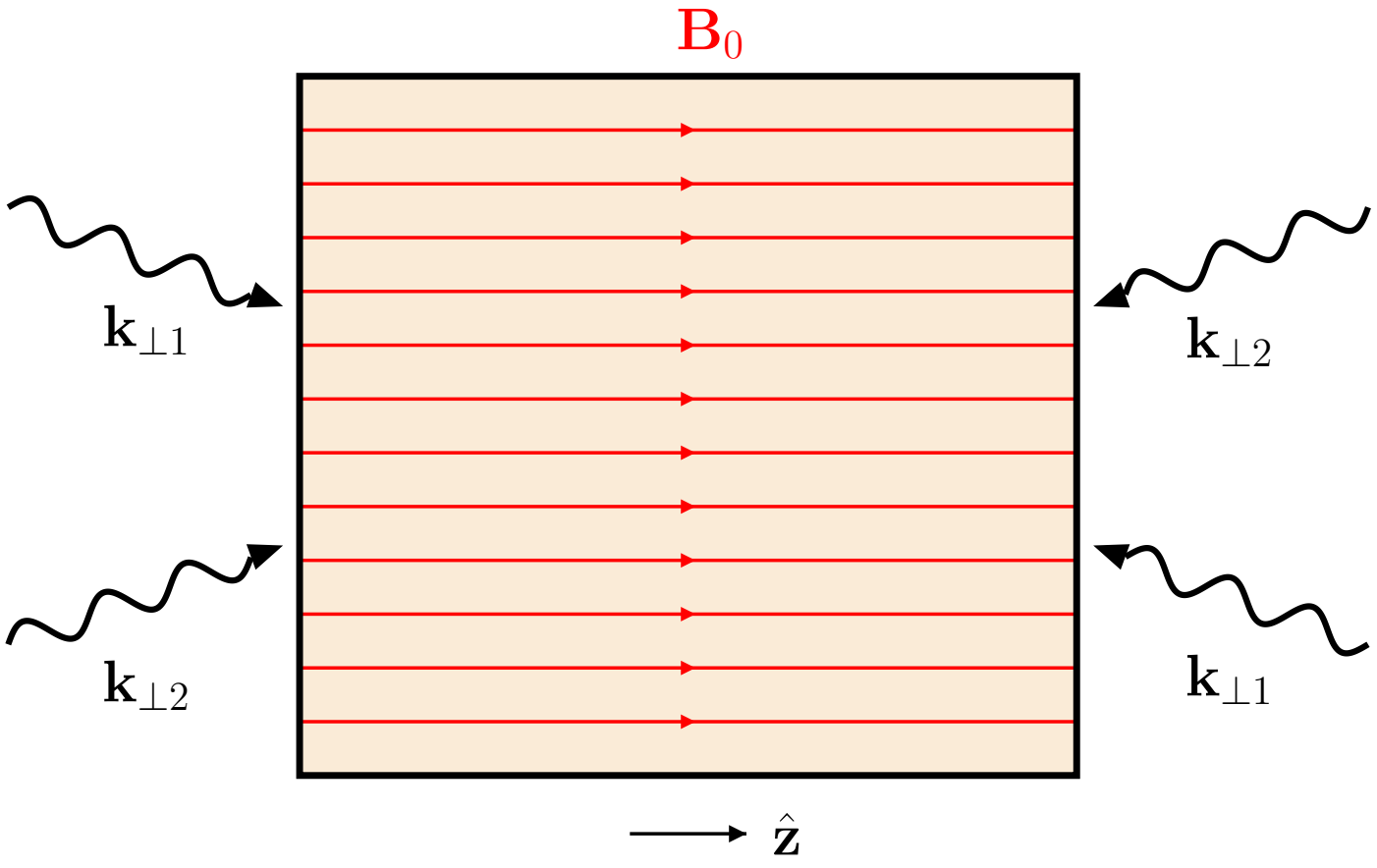

By analogy, in our homogeneous plasma, we will coherently drive counterpropagating SAWs of a particular frequency $\omega_\mathrm{Ad}$ and hence $|k_{\parallel \mathrm{d}}|$, but with parallel wavenumbers of opposite sign and with two different choices of~$\mathbf{k}_\perp$. Crucially, this allows the waves to interact, as SAWs of the same polarisation, i.e., with $\mathbf{k}_{\perp 1} \propto \mathbf{k}_{\perp 2}$, would have a vanishing nonlinear coupling. This is a consequence of the structure of the MHD nonlinearity \mycitep[see, e.g.,][]{howes_alfven_2013}. In our model, these waves are excited by an external antenna inspired by \mycitet{tenbarge_oscillating_2014}: we add a term in the equations corresponding to a free current $j_{z, \mathrm{ant}} (\mathbf{r},t) = - c \nabla^2 A_{z, \mathrm{ant}}(\mathbf{r},t)/4 \pi$, and hence an additional vector potential (we use the Coulomb gauge). Overall, the system (caricatured in \cref{fig:diagram}) is designed so that SAW `pump waves' of the same frequency but different $\mathbf{k}_\perp$ can propagate through each other and interact nonlinearly. The antenna will be tuned to ensure that they saturate at small amplitudes, as appropriate for TAEs in real devices, and also to enforce the weak-turbulence limit for the pump waves. The key question is then how large the `zonal' modes ($k_\parallel = 0$ and $\omega = 0$) become as a function of the pump-wave amplitude. Answering it will help us estimate the strength of the ensuing shearing effect on the underlying microturbulence, and thus of any thermal transport reduction in magnetic fusion devices.

\subsection{Derivation of forcing} \label{sec:antderiv}

To complete the governing equations of our model, it remains to connect $F^\pm$ in \cref{eq:rmhdzeta} with the antenna described in \S\cref{sec:antmotiv}. The Alfvénic term $v_\mathrm{A} \partial_z \nabla_\perp^2 \Psi$ in \cref{eq:lapphi} originates from the curl of the $\mathbf{j} \times \mathbf{B}_0$ force,

\begin{equation}
    \hat{\mathbf{z}} \cdot \left[ \nabla \times \left( \mathbf{j} \times \mathbf{B}_0 \right) \right] = B_0 \partial_z j_z,
\end{equation} where we have used $\nabla \cdot \mathbf{B}_0 = 0$, $\nabla \cdot \mathbf{j} = 0$ and $\nabla \mathbf{B}_0 =0$. Introducing an antenna current $\mathbf{j}_{\mathrm{ant}}$ amounts to replacing $\mathbf{j}$ with $\mathbf{j} - \mathbf{j}_{\mathrm{ant}}$, so we add a new linear term on the right-hand side of \cref{eq:lapphi} in opposition to $v_\mathrm{A} \partial_z \nabla^2_\perp \Psi$. Then, in \cref{eq:Psieq} and \cref{eq:lapphi}, we identify $F_\Psi = 0$ (because, in ideal MHD, the antenna current does not enter explicitly in the induction equation) and $\nabla_\perp^2 F_\Phi = - v_\mathrm{A} \partial_z \nabla_\perp^2 \Psi_{\mathrm{ant}} (\mathbf{r}, t)$. This implies that the two fields $\zeta^\pm$ are forced equally (a `balanced drive'): in the Elsasser formulation the antenna becomes

\begin{equation} \label{eq:forcingterm}
    F^\pm = F = -v_\mathrm{A} \partial_z \Psi_{\mathrm{ant}} (\mathbf{r}, t).
\end{equation}

Finally, the spatial and temporal dependence of the antenna potential $\Psi_{\mathrm{ant}}$ must be specified. Defining $\Psi_\mathrm{A}$ as a constant amplitude, we set

\begin{equation}
    \Psi_{\mathrm{ant}} (\mathbf{r}, t) = \Psi_\mathrm{A} M(\mathbf{r}) \sin(\omega_\mathrm{Ad} t), \quad
    M(\mathbf{r}) = \sum_i \exp( i \mathbf{k}_{\mathrm{A}i} \cdot \mathbf{r}),
\end{equation} where $\mathbf{k}_{\mathrm{A}i}$ are the wavevectors of the Alfvén waves that we wish to excite with frequency~$\omega_\mathrm{Ad}$. They are all driven with equal strength. In our simulations, we will excite four wavevectors with the same $|k_z| = k_{\parallel \mathrm{d}} = |\omega_\mathrm{Ad}|/v_\mathrm{A}$. This will allow us to make two arbitrary choices $\mathbf{k}_{\perp 1}$ and $\mathbf{k}_{\perp 2}$ for $\mathbf{k}_{\mathrm{A}1}$ and $\mathbf{k}_{\mathrm{A}2}$, while also exciting additional waves with $\mathbf{k}_{\mathrm{A}3} = -\mathbf{k}_{\mathrm{A}1}$ and $\mathbf{k}_{\mathrm{A}4} = -\mathbf{k}_{\mathrm{A}2}$ to make $\Psi_\mathrm{ant}(\mathbf{r},t)$ a real field. The forcing term \cref{eq:forcingterm} will generate waves travelling both up and down the guide field for each of these~$\mathbf{k}_{\mathrm{A}i}$.

\subsection{Frequency doubling}  \label{sec:freqdoubl}

Before deriving a further reduced model for our system, it is worth briefly investigating why SAWs were not thought to produce `zonal' modes in a homogeneous MHD slab. The condition that only counterpropagating SAWs can interact implies that modes created directly with $k_\parallel=0$ should be at double the frequency of the pump wave instead of $\omega = 0$. This can be seen as follows. Consider a single `collision' between two counterpropagating wavepackets, in a system where the nonlinearity is weak. As their interaction is small, the wavepackets are hardly modified as they pass each other, so can be thought of as quasiparticles. Then the resonance conditions (corresponding to momentum and energy conservation) for a three-wave interaction require \mycitep[see, e.g.,][]{shebalin_anisotropy_1983, schekochihin_mhd_2022}:

\begin{equation} \label{eq:qpmom}
    \mathbf{p} + \mathbf{q} = \mathbf{k}, \quad
    \omega_\mathbf{p} + \omega_\mathbf{q} = \omega_\mathbf{k},
\end{equation} where $\mathbf{p}, \mathbf{q},$ and $\mathbf{k}$ are the three wavevectors with associated frequencies $\omega_\mathbf{p}, \omega_\mathbf{q},$ and~$\omega_\mathbf{k}$. To generate $k_\parallel = 0$, we must have $p_\parallel = -q_\parallel$ — and as the wavepackets must be counterpropagating to interact, the resulting frequency $\omega_\mathbf{k} = \omega_\mathbf{p}^\pm + \omega_\mathbf{q}^\mp = \mp p_\parallel v_\mathrm{A} \pm q_\parallel v_\mathrm{A} = \mp 2 p_\parallel v_\mathrm{A}$ is doubled. It would appear that the desired low-frequency modes are not present. However, we will show that over longer timescales, an additional component at $\omega=0$ builds up, and saturates at a magnitude similar to the double-frequency modes.  

\section{Truncated system} \label{sec:analmethods}

In this section, we reduce the model of \S\cref{sec:antennarMHD} to distil the essential physics of the problem. Equations \cref{eq:rmhdzeta} contain two timescales: the linear timescale $\tau_\mathrm{A}$ and the nonlinear timescale $\tau_{\mathrm{nl}}$. To expose the dynamics of interest — the nonlinear evolution of the modes — it is beneficial to eliminate the `fast' linear oscillations. This is done in \S\cref{sec:freqexp}, following a truncation of the system in parallel wavenumber in \S\cref{sec:paralleltrunc}. Both of these steps are enabled by the weak-turbulence condition $\tau_\mathrm{A} \ll \tau_{\mathrm{nl}}$. The result is a closed set of equations with a small number of degrees of freedom, which will then be suitable for further analysis in~\S\cref{sec:analresults}.

\subsection{Truncation of parallel wavenumbers} \label{sec:paralleltrunc}

To help justify our truncation of parallel wavenumbers, and also as a first step to separating $\tau_\mathrm{A}$ and $\tau_\mathrm{nl}$, we eliminate the linear term in \cref{eq:rmhdzeta} via a parallel Fourier decomposition, and by factoring out $\exp(-i \omega_\mathrm{l}^\pm t)$ from each Fourier component:

\begin{equation} \label{eq:linshift}
    \zeta^\pm (t, \mathbf{r}) = \sum_{k_\parallel} \zeta_{k_\parallel}^\pm (t, \mathbf{r}_\perp) e^{-i \omega_\mathrm{l}^\pm t + i k_\parallel z} =\sum_{k_\parallel} \zeta_{k_\parallel}^\pm (t, \mathbf{r}_\perp) e^{\pm ik_\parallel v_\mathrm{A} t + i k_\parallel z},
\end{equation}

\begin{equation} \label{eq:drivetransform}
    F (t, \mathbf{r}) = \sum_{k_\parallel} F_{k_\parallel} (t, \mathbf{r}_\perp) e^{i k_\parallel z}.
\end{equation} Substituting \cref{eq:linshift} and \cref{eq:drivetransform} into \cref{eq:rmhdzeta} (with $k_\parallel$ relabelled as $k_\parallel'$), multiplying by $e^{\mp i k_\parallel v_\mathrm{A} t - i k_\parallel z}$, and then integrating over the parallel domain, we find

\begin{equation} \label{eq:Schekgen}
    \partial_t \zeta_{k_\parallel}^\pm = \sum_{k'_\parallel} N[\zeta^\mp_{k'_\parallel}, \zeta^\pm_{k_\parallel - k'_\parallel}] e^{\mp i 2 k'_\parallel v_\mathrm{A} t} + F_{k_\parallel} e^{\mp i k_\parallel v_\mathrm{A} t}.
\end{equation} By separating out the $k_\parallel = 0$ `condensate' modes explicitly, we get the following equations (cf. \mycitet{schekochihin_weak_2012}):

\begin{equation} \label{eq:Schekcond}
    \partial_t \zeta^\pm_0 = N[\zeta_0^\mp, \zeta_0^\pm] + \sum_{k_\parallel \neq 0} N[\zeta_{k_\parallel}^\mp, \zeta_{-k_\parallel}^\pm] e^{\mp i2k_\parallel v_\mathrm{A} t},
\end{equation}

\begin{equation} \label{eq:Schekwave}
    \partial_t \zeta^\pm_{k_\parallel} = N[\zeta_{0}^\mp, \zeta_{k_\parallel}^\pm] + N[\zeta_{k_\parallel}^\mp, \zeta_{0}^\pm] e^{\mp i2k_\parallel v_\mathrm{A} t} + \sum_{k_\parallel' \neq 0, k_\parallel } N[\zeta_{k_\parallel'}^\mp, \zeta_{k_\parallel-k_\parallel'}^\pm] e^{\mp i 2 k_\parallel' v_\mathrm{A} t} + F_{k_\parallel} e^{\mp i k_\parallel v_\mathrm{A} t}.
\end{equation}

We note that the condensate is a subsystem with no linear physics, containing only self-advection and an effective forcing modulated at high frequency. At finite $k_\parallel$, in \cref{eq:Schekwave}, the dominant nonlinear physics is the advection by the condensate, with the fast oscillation of the other terms greatly reducing their effect\footnote{This can be seen more simply by solving the equations of \cref{eq:qpmom} simultaneously. It is not possible for all three modes to obey the linear dispersion relation with $k_\parallel \neq 0$ — asymptotically all nonlinear interactions in weak MHD turbulence must involve a condensate mode, and thus preserve the parallel wavenumber.}. The forcing term will later be shown to contain a constant part for $|k_\parallel| / k_{\parallel \mathrm{d}} = 1$, as required for a resonant drive by the antenna.

\begin{figure} 
    \centering
    \includegraphics[width=0.6\linewidth]{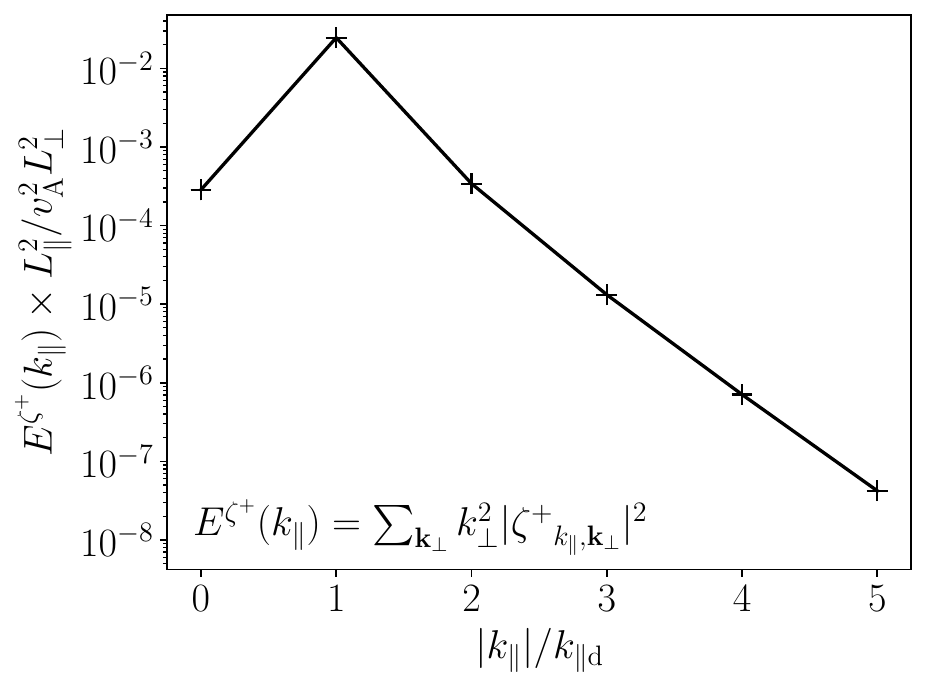}
    \caption{The saturated energy content of different parallel wavenumbers for an antenna of strength $\Psi_\mathrm{A} = 2 \times 10^{-4} v_\mathrm{A} L_\perp^2 / L_\parallel$. The data is derived from a simulation of the kind described in \S\cref{sec:numver}. In the ensuing numerical and analytical work, modes beyond $|k_\parallel| / k_{\parallel \mathrm{d}}=2$ will be removed. This is a good approximation in this case (the `strongest' antenna considered in the later scans described in \S\cref{sec:numver}), and better still for the weaker antennae.}
    \label{fig:par_spec}
\end{figure}

By considering \cref{eq:Schekcond} as a stochastic equation, or via dimensional analysis \mycitep{schekochihin_weak_2012}, we can estimate the saturated amplitudes of the condensate modes as a function of the driven modes: denoting $\zeta^\pm_{n k_{\parallel \mathrm{d}}}$ as $\zeta^\pm_n$,

\begin{equation} \label{eq:schekscale}
	\zeta_0^\pm \sim \omega_\mathrm{Ad}^{-1} k_\perp^2 \zeta_1^+ \zeta_1^-.
\end{equation} As the system has a balanced forcing, we assume $\zeta_n^+ \sim \zeta_n^-$ (which is indeed the case in our simulations), allowing us to define a small parameter

\begin{equation} \label{eq:epsilondefn}
	\epsilon \equiv \left( \frac{\tau_\mathrm{A}}{\tau_\mathrm{nl}} \right)_{k_\parallel = k_{\parallel \mathrm{d}}} = \frac{k_\perp^2 \zeta_1^\pm}{\omega_\mathrm{Ad}} \ll 1.
\end{equation} Then \cref{eq:schekscale} implies $\zeta_0^\pm \sim \epsilon \zeta_1^\pm$. Using \cref{eq:Schekgen}, we find more generally

\begin{equation}
    \frac{\zeta_n^\pm}{\zeta^\pm_1} \sim \epsilon^{\min\{ |n-1|, |n+1| \}},
\end{equation} so $\zeta_{\pm 2}^\pm / \zeta_1^\pm \sim \zeta_{0}^\pm/\zeta_{1}^\pm \sim \epsilon$, and all other components are higher order. The result is a steep energy spectrum in $k_\parallel$ (which can indeed be seen in the simulations described in \S\cref{sec:numver} — see \cref{fig:par_spec}), allowing us to retain only $|k_\parallel| / k_{\parallel \mathrm{d}} = 0,1,2$ at all times. This truncation should be valid both transiently and in statistical steady state, provided the weak-turbulence condition \cref{eq:epsilondefn} is well satisfied.

\subsection{Non-perturbative frequency expansion} \label{sec:freqexp}

\begin{figure} 
	\centering
	\includegraphics[width = 0.48\textwidth]{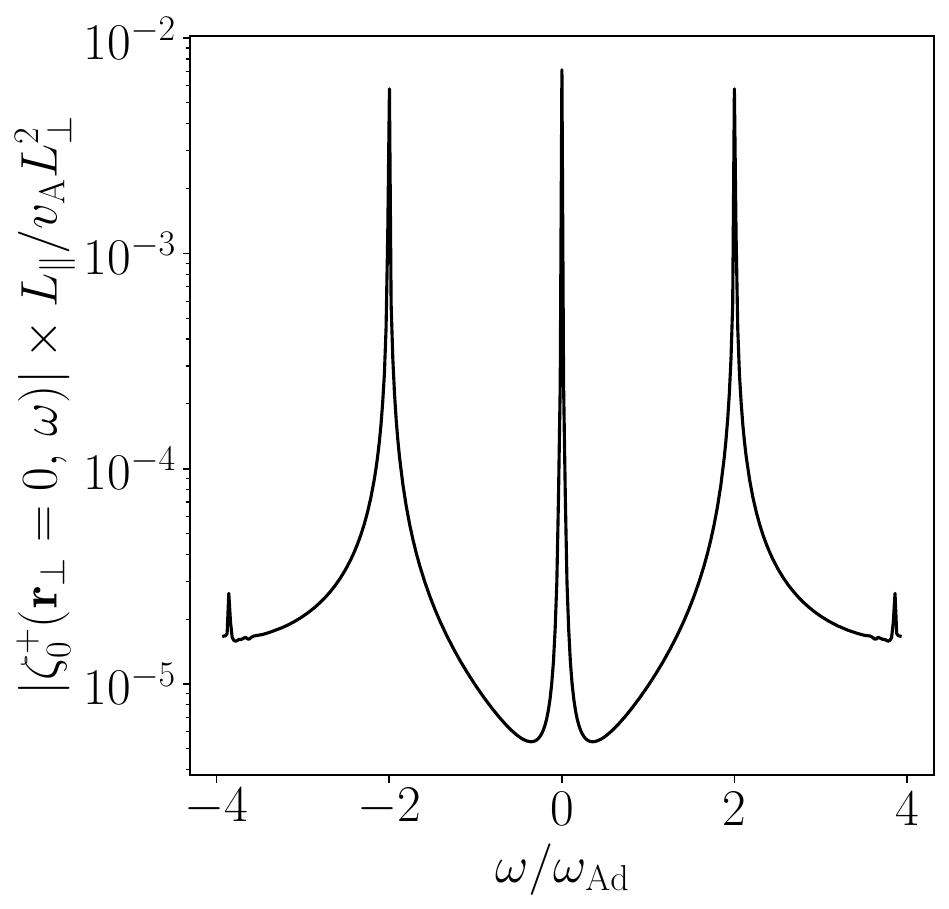}
	\includegraphics[width = 0.48\textwidth]{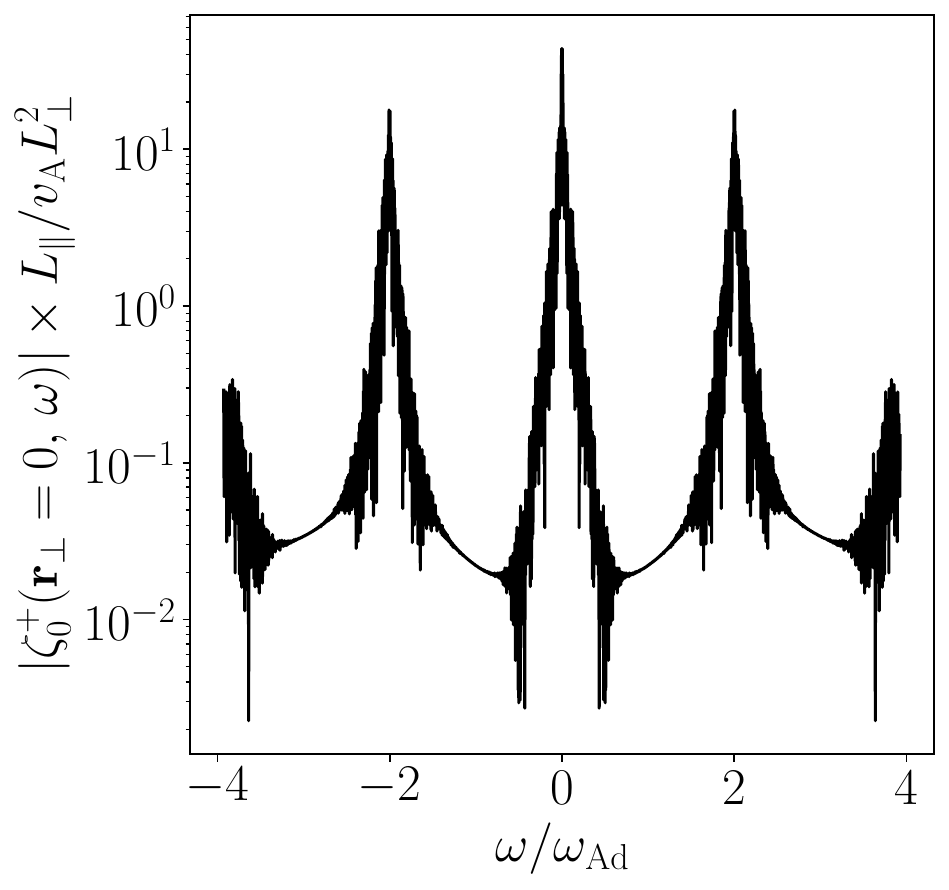}
	\caption{The frequency structure of the $k_\parallel = 0$ condensate for $\Psi_\mathrm{A} = 2 \times 10^{-7} v_\mathrm{A} L_\perp^2 / L_\parallel$ (left) and $\Psi_\mathrm{A} = 2 \times 10^{-3} v_\mathrm{A} L_\perp^2 / L_\parallel$ (right), from simulations of the type described in \S\cref{sec:numver} (although the parameter scan shown there does not extend as far as the strongly driven case displayed here). The only significant difference is the relative width of the peaks, corresponding to the ratio~$\tau_\mathrm{A}/\tau_{\mathrm{nl}}$. Frequency plots do not change significantly when evaluated at different $\mathbf{r}_\perp$.}
	\label{fig:perturbative}
\end{figure}

In many systems, the effect of a small nonlinearity is to give a finite width to the linear peak in frequency space. This can be interpreted as an effective quality factor, as nonlinearities remove energy from the mode in a manner somewhat analogous to conventional damping. This is also the prescription of time-dependent perturbation theory in quantum mechanics. If this were the case in weak RMHD turbulence, then the rewriting described by \cref{eq:linshift} would leave $\zeta^\pm_{k_\parallel}$ containing only slow time dependence. However, all wave modes are forced to interact with the condensate, which in this coherently driven system has characteristic frequencies $\pm 2 \omega_\mathrm{Ad}$. We make an ansatz that frequency sidebands may be generated, and hence $\zeta^\pm_{k_\parallel}$ may still contain fast time dependence. The fields will contain several narrow peaks, with the ratio of the peak width to the peak separation of order~$\tau_\mathrm{A} / \tau_{\mathrm{nl}}$, as shown in \cref{fig:perturbative}. Let us separate the slow and fast time variation (which, as per \cref{eq:linshift}, are defined with respect to the linear frequency of the mode):

\begin{equation}
    \zeta^\pm_n = \bar{\zeta}_n^\pm + \tilde{\zeta}_n^\pm,
\end{equation} where, to isolate the slowly varying part of the field, we use a time average over an intermediate time $\tau_\mathrm{A} \ll T \ll \tau_\mathrm{nl}$:

\begin{equation}
    \bar{\zeta}^\pm_n (t) = \frac{1}{T}\int_{t}^{t+T} \mathrm{d}t' \zeta_n^\pm (t').
\end{equation}

For the fields at the retained parallel wavenumbers ($n = 0, \pm 1, \pm2$), we make a series of substitutions. In the $n=0$ condensate, we allow for narrow peaks
at $\omega/\omega_\mathrm{Ad} = 0, \pm 2$. As $\zeta_0^\pm$ is real, this can be expressed as

\begin{equation} \label{eq:zeta0}
    \begin{split}
\zeta^\pm_0 &= \bar{\zeta}_0^\pm + \tilde{\zeta}_0^\pm \approx \bar{\zeta}_0^\pm + \xi_0^\pm e^{\mp i2 \omega_\mathrm{Ad} t} + (\xi_0^\pm)^\star e^{\pm i2 \omega_\mathrm{Ad} t},
\end{split}
\end{equation} where the two-dimensional fields $\bar{\zeta}^\pm_0(\mathbf{r}_\perp)$ and $\xi^\pm_0(\mathbf{r}_\perp)$ vary slowly in time at the nonlinear frequency. For $k_\parallel = \pm k_{\parallel \mathrm{d}}$, we expect any sidebands to be negligible compared to the driven peaks, so we simply write 

\begin{equation} \label{eq:zeta1}
    \zeta_1^\pm \approx \bar{\zeta}_1^\pm.
\end{equation} Note that $\zeta_{-1}^\pm = (\zeta_{1}^\pm)^\star$ by Hermitian symmetry. Finally, for $k_\parallel = \pm 2 k_{\parallel \mathrm{d}}$, we allow for an `off-shell' component\footnote{We refer to the fields $\xi^\pm$ as `off-shell' because their frequencies do not (even approximately) satisfy the linear dispersion relation $\omega_\mathrm{l}^\pm = \mp k_\parallel v_\mathrm{A}$.} $\xi^\pm_2$ at overall zero frequency, but neglect any possible dynamics at~$\pm 4 \omega_\mathrm{Ad}$, which are much faster than the characteristic timescales of the system:

\begin{equation} \label{eq:zeta2}
    \begin{split}
\zeta_2^{\pm} & = \bar{\zeta}_2^\pm + \tilde{\zeta}_2^\pm \approx \bar{\zeta}_2^\pm + \xi_2^\pm e^{\mp i2 \omega_\mathrm{Ad} t}.
\end{split}
\end{equation} For numerical evidence that these decompositions capture the frequency structure of our fields, see \cref{fig:freq_structure}. \Cref{fig:freq_schematic} shows a schematic of the peaks, with the on- and off-shell components labelled individually.

\begin{figure} 
  \centering
  \includegraphics[width = 0.48\textwidth]{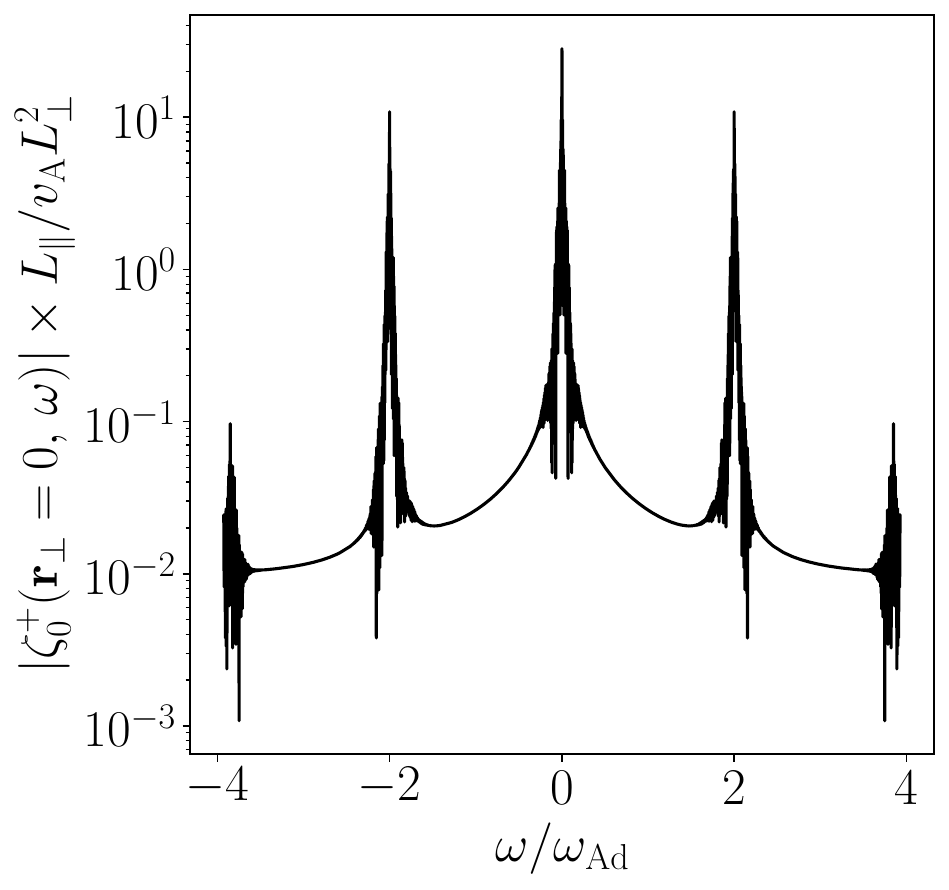}
  \includegraphics[width = 0.48\textwidth]{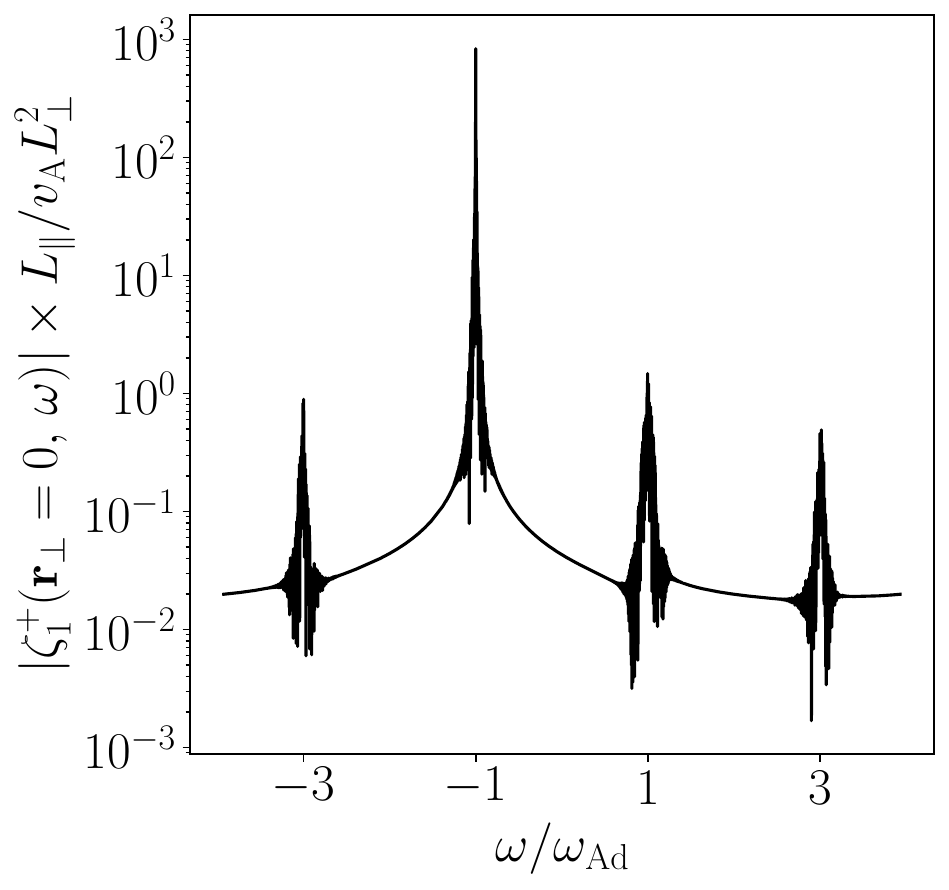}
  \includegraphics[width = 0.48\textwidth]{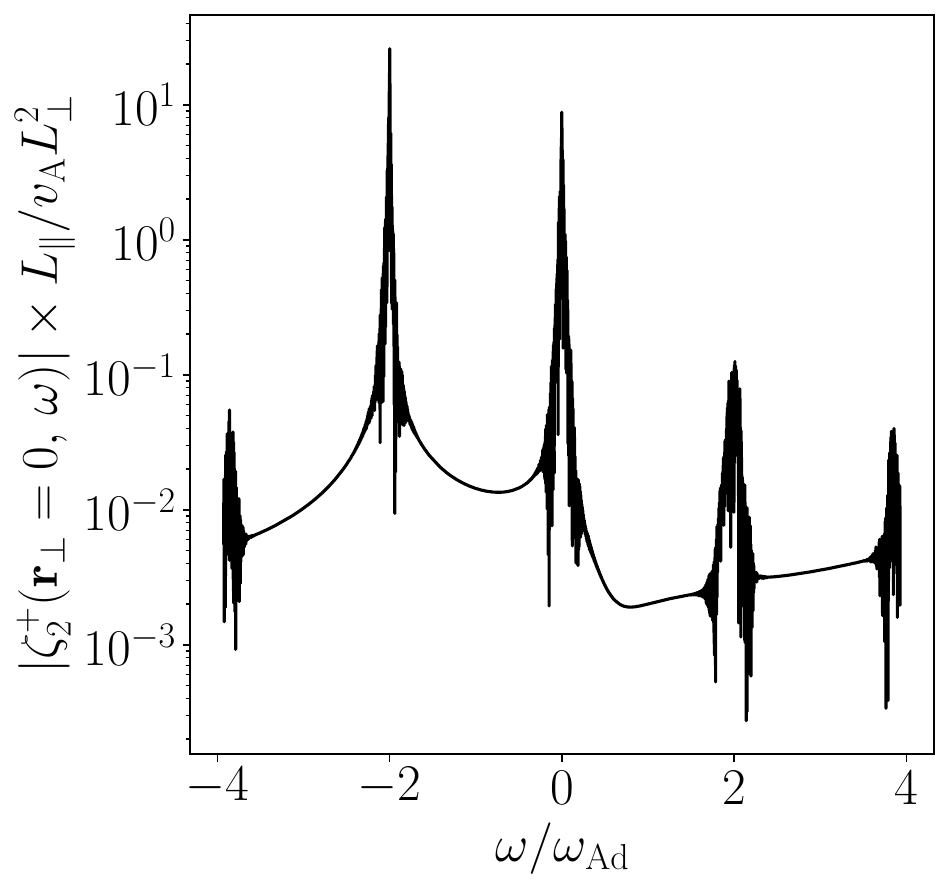}
  \caption{The frequency structure of the field $\zeta^+_{k_\parallel}$ for $k_\parallel/k_{\parallel \mathrm{d}} = 0,1,2$ and $\Psi_\mathrm{A} = 2 \times 10^{-4} v_\mathrm{A} L_\perp^2 / L_\parallel$, taken from the simulation described in \S\cref{sec:numver}. The plots have been recentred to show absolute frequency, so the fields are not shifted relative to the linear frequency as in \cref{eq:linshift}.}
\label{fig:freq_structure}
\end{figure}
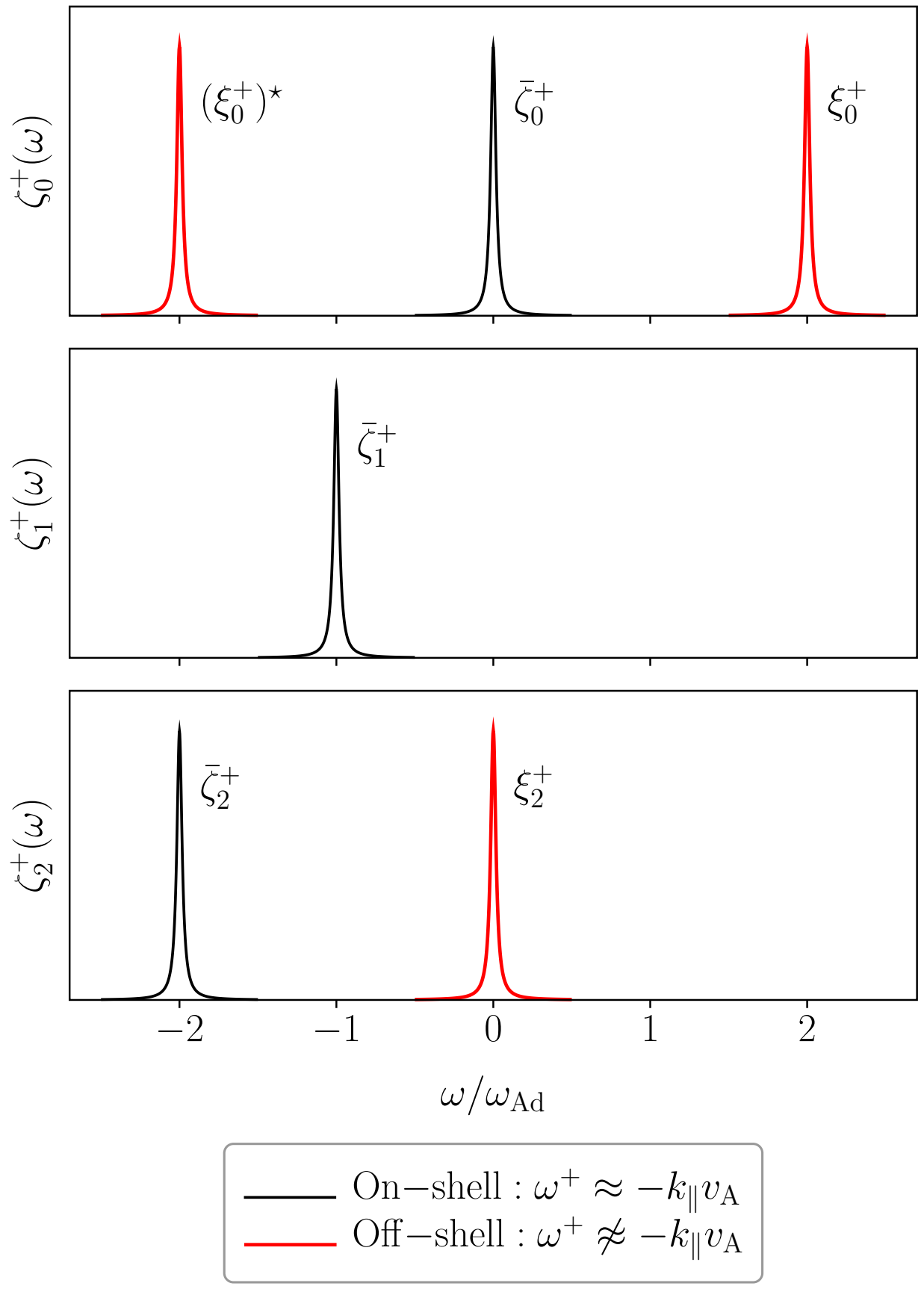
\begin{figure}
	\centering
		\resizebox{0.6\textwidth}{!}{%
			\begin{tikzpicture}
				\pgfplotsset{
					spectrum axis/.style={
						scale only axis,
						width=5.104100in,
						height=1.862417in,
						xmin=-2.7, xmax=2.7,
						ymin=0, ymax=1.15,
						xtick={-2,-1,0,1,2},
						ytick=\empty,
						axis lines=box,
						axis line style={line width=0.803pt},
						tick align=outside,
						tick style={black,line width=0.803pt},
						major tick length=0.048611in,
						xtick pos=bottom,
						tick label style={font=\sffamily\fontsize{17.28}{20.736}\selectfont},
						label style={font=\sffamily\fontsize{17.28}{20.736}\selectfont},
						xlabel style={at={(axis description cs:0.5,-0.22)}, anchor=north},
						axis on top,
						clip=false,
					},
					on shell/.style={samples=501, no marks, black, line width=1.2045pt},
					off shell/.style={samples=501, no marks, red, line width=1.505625pt},
				}
				
				\begin{groupplot}[
					group style={group size=1 by 3, vertical sep=0.198611in},
					spectrum axis,
					]
					\nextgroupplot[xticklabels=\empty]
					\addplot[on shell, domain=-0.5:0.5]
					{0.0004/((x-0)^2+0.0004)};
					\addplot[off shell, domain=-2.5:-1.5]
					{0.0004/((x+2)^2+0.0004)};
					\addplot[off shell, domain=1.5:2.5]
					{0.0004/((x-2)^2+0.0004)};
					\node[anchor=north west,font=\sffamily\fontsize{17.28}{20.736}\selectfont]
					at (axis cs:-1.92,0.9) {$(\xi^+_0)^\star$};
					\node[anchor=north west,font=\sffamily\fontsize{17.28}{20.736}\selectfont]
					at (axis cs:0.08,0.9) {$\bar{\zeta}^+_0$};
					\node[anchor=north west,font=\sffamily\fontsize{17.28}{20.736}\selectfont]
					at (axis cs:2.08,0.9) {$\xi^+_0$};
					
					\nextgroupplot[xticklabels=\empty]
					\addplot[on shell, domain=-1.5:-0.5]
					{0.0004/((x+1)^2+0.0004)};
					\node[anchor=north west,font=\sffamily\fontsize{17.28}{20.736}\selectfont]
					at (axis cs:-0.92,0.9) {$\bar{\zeta}^+_1$};
					
					\nextgroupplot[xlabel={$\omega / \omega_{\mathrm{Ad}}$}]
					\addplot[on shell, domain=-2.5:-1.5]
					{0.0004/((x+2)^2+0.0004)};
					\addplot[off shell, domain=-0.5:0.5]
					{0.0004/(x^2+0.0004)};
					\node[anchor=north west,font=\sffamily\fontsize{17.28}{20.736}\selectfont]
					at (axis cs:-1.92,0.9) {$\bar{\zeta}^+_2$};
					\node[anchor=north west,font=\sffamily\fontsize{17.28}{20.736}\selectfont]
					at (axis cs:0.08,0.9) {$\xi^+_2$};
				\end{groupplot}
				
				\node[rotate=90,anchor=center,font=\sffamily\fontsize{17.28}{20.736}\selectfont]
				at ([xshift=-0.23in]group c1r1.west) {$\zeta_0^+(\omega)$};
				\node[rotate=90,anchor=center,font=\sffamily\fontsize{17.28}{20.736}\selectfont]
				at ([xshift=-0.23in]group c1r2.west) {$\zeta_1^+(\omega)$};
				\node[rotate=90,anchor=center,font=\sffamily\fontsize{17.28}{20.736}\selectfont]
				at ([xshift=-0.23in]group c1r3.west) {$\zeta_2^+(\omega)$};
				
				\node[
				draw=gray!80,
				fill=white,
				fill opacity=0.8,
				text opacity=1,
				rounded corners=4pt,
				line width=1.00375pt,
				below=0.86in of group c1r3.south,
				anchor=north,
				font=\sffamily\fontsize{17.28}{20.736}\selectfont,
				inner xsep=0.11in,
				inner ysep=0.13in,
				] {
					\begin{tabular}{@{}ll@{}}
						\tikz{\draw[black,line width=1.2045pt] (0,0) -- (0.556in,0);} &
						$\mathrm{On}\mathrm{-shell} : \omega^+ \approx - k_\parallel v_{\mathrm{A}}$ \\
						\tikz{\draw[red,line width=1.505625pt] (0,0) -- (0.556in,0);} &
						$\mathrm{Off}\mathrm{-shell} : \omega^+ \not \approx - k_\parallel v_{\mathrm{A}}$
					\end{tabular}
				};
			\end{tikzpicture}%
		}
	\caption{A schematic of the frequency structure of the $\zeta^+$ fields that is assumed in the analytical expansion described in \S\cref{sec:freqexp}. Like \cref{fig:freq_structure}, the plots have been recentred to show absolute frequency, rather than being relative to the linear frequency as in \cref{eq:linshift}.}
	\label{fig:freq_schematic}
\end{figure}

Finally, we substitute \crefrange{eq:zeta0}{eq:zeta2} into \cref{eq:Schekcond} and \cref{eq:Schekwave}, and separate the dynamics at multiples of $\omega_\mathrm{Ad}$ — each forbidden from cross-talk with the others by the small size of~$\tau_\mathrm{A}/\tau_{\mathrm{nl}}$. At each frequency, only the leading-order contributions in $\epsilon$ are retained. We also use $F_{1} e^{\mp i \omega_\mathrm{Ad} t} = \mp D \left[1 - e^{\mp i 2 \omega_\mathrm{Ad} t}\right]$, with $D (\mathbf{r}_\perp) = \omega_\mathrm{Ad} \Psi_\mathrm{A} M_\perp (\mathbf{r}_\perp) / 2$, and $M_\perp(\mathbf{r}_\perp) = \sum_{i, k_\parallel = k_{\parallel d}} \exp( i \mathbf{k}_{\perp \mathrm{A}i} \cdot \mathbf{r}_\perp)$. We find that the on-shell fields evolve dynamically on slow, nonlinear timescales as

\begin{align}
			\partial_t \bar{\zeta}_0^\pm &= N[\bar{\zeta}_0^\mp,\bar{\zeta}_0^\pm] + \left(  N[\xi_0^\mp, \xi_0^\pm] + N[\xi_2^\mp, (\xi_2^\pm)^\star] + \mathrm{c.c.} \right), \label{eq:barzetazerofirst} \\
			\partial_t \bar{\zeta}_1^\pm &= N[\bar{\zeta}_0^\mp,\bar{\zeta}_1^\pm] + N[\bar{\zeta}_1^\mp, (\xi_0^\pm)^\star] + N[(\bar{\zeta}_1^\mp)^\star, \xi_2^\pm] \mp D, \label{eq:barzeta1first} \\
		\partial_t \bar{\zeta}_2^\pm & = N[\bar{\zeta}_0^\mp,\bar{\zeta}_2^\pm]  + N[\xi_0^\mp, \xi_2^\pm] + N[\xi_2^\mp, (\xi_0^\pm)^\star], \label{eq:barzeta2first}
\end{align} while the off-shell field amplitudes are fixed instantaneously by the pump waves

\begin{align}
		\mp i 2 \omega_\mathrm{Ad} \xi_0^\pm 
		&= N[\bar{\zeta}_1^\mp,(\bar{\zeta}_1^\pm)^\star],  \label{eq:xi0eqn} \\
		\mp i 2 \omega_\mathrm{Ad} \xi_2^\pm 
		& = N[\bar{\zeta}_1^\mp,\bar{\zeta}_1^\pm].  \label{eq:xi2eqn} \\
\end{align} Hence $\xi_0^\pm$ and $\xi_2^\pm$ — directly driven by the nonlinear interaction of the counterpropagating pump modes — do not evolve in time independently, but are slaved to the amplitudes of~$\bar{\zeta}_1^\pm$. Substituting \cref{eq:xi0eqn} and \cref{eq:xi2eqn} into \crefrange{eq:barzetazerofirst}{eq:barzeta2first}, we obtain the final closed equations for the on-shell fields:

\begin{align} 
\partial_t \bar{\zeta}_0^\pm = &\,N[\bar{\zeta}_0^\mp,\bar{\zeta}_0^\pm] \notag \\
&+ \frac{1}{4\omega^2_{\mathrm{Ad}}} N\left[ N[\bar{\zeta}_1^\pm,(\bar{\zeta}_1^\mp)^\star],N[\bar{\zeta}_1^\mp,(\bar{\zeta}_1^\pm)^\star]\right] + \mathrm{c.c.} \notag \\
& - \frac{1}{4 \omega^2_{\mathrm{Ad}}} N\left[ N[\bar{\zeta}_1^\pm,\bar{\zeta}_1^\mp] , N[(\bar{\zeta}_1^\mp)^\star,(\bar{\zeta}_1^\pm)^\star]\right] + \mathrm{c.c.}, \label{eq:barzetazero} \displaybreak[0] \\
\partial_t \bar{\zeta}_1^\pm =& \, N[\bar{\zeta}_0^\mp,\bar{\zeta}_1^\pm] \notag \\
& \mp \frac{i}{2 \omega_\mathrm{Ad}} N\left[\bar{\zeta}_1^\mp,N[(\bar{\zeta}_1^\mp)^\star, \bar{\zeta}_1^\pm] \right] \notag \\
& \pm \frac{i}{2 \omega_\mathrm{Ad}} N\left[(\bar{\zeta}_1^\mp)^\star, N[\bar{\zeta}_1^\mp,\bar{\zeta}_1^\pm]\right] \notag \\ 
& \mp D,  \label{eq:barzeta1} \displaybreak[0] \\
\partial_t \bar{\zeta}_2^\pm  = &\,N[\bar{\zeta}_0^\mp,\bar{\zeta}_2^\pm]\notag \\
& + \frac{1}{4 \omega^2_{\mathrm{Ad}}} N\left[ N[\bar{\zeta}_1^\pm,(\bar{\zeta}_1^\mp)^\star],N[\bar{\zeta}_1^\mp,\bar{\zeta}_1^\pm]\right] \notag \\
& - \frac{1}{4 \omega^2_{\mathrm{Ad}}}N\left[ 
N[\bar{\zeta}_1^\pm,\bar{\zeta}_1^\mp],N[(\bar{\zeta}_1^\mp)^\star, \bar{\zeta}_1^\pm]\right].  \label{eq:barzeta2}
\end{align} If desired, these equations could be simulated directly and efficiently, as large time steps are allowed for the slow nonlinear dynamics. Note that, although $\bar{\zeta}_2^\pm$ does not enter the dynamical equations \cref{eq:barzetazero} or \cref{eq:barzeta1}, if the $\zeta_{2}^\pm$ fields had been ignored entirely\footnote{This might have appeared to be a reasonable first assumption, if argued that the only important physics is the transfer of energy from the driven wavenumber into the condensate, which then acts back on the driving.}, the driving terms due to $\xi^\pm_2$ would have been missed. If more parallel wavenumbers are retained, they do not enter \crefrange{eq:barzetazero}{eq:barzeta2} at leading order.

In summary, in this section we have obtained a closed set of equations with a reduced number of degrees of freedom, and only nonlinear time evolution. This will facilitate the ensuing analysis and discussion in \S\cref{sec:analresults}.

\section{Saturated amplitudes} \label{sec:analresults}

Here we use the reduced model derived in \S\cref{sec:analmethods} to obtain approximate scalings for the saturated amplitudes of the different frequency components. This is done assuming that the fields do not acquire specific phases that conspire to change the dynamics significantly; this assumption is verified numerically in \S\cref{sec:numver}. The transient behaviour before saturation is discussed in \cref{app:transient}. We de-emphasise it here because the antenna is unrealistic, and because we wish to focus on the saturated, statistical steady state of the system. In~\S\cref{sec:lit}, the relation between the `zonal' modes and the pump waves will be compared to other predictions appearing in the literature. The exact mechanism of their generation will be discussed in detail in \S\cref{sec:mech}.

Using \crefrange{eq:barzetazero}{eq:barzeta2}, approximate scalings can be found for the relative size of the fields in the saturated state. By assuming that the perpendicular Fourier transforms of the 2D fields $\bar{\zeta}_n(\mathbf{r}_\perp)$ and ${\xi}_n(\mathbf{r}_\perp)$ are peaked around $k_\perp = k_{\perp \mathrm{A}}$\footnote{In the present study, we consider only box-scale forcing, for which this is found to be true. However, we note that \cref{eq:barzetazero} takes the form of driven two-dimensional MHD, viz. $\partial_t \bar{\zeta}^\pm_0 = N[ \bar{\zeta}^\mp_0, \bar{\zeta}^\pm_0] + F^\pm_{2D}$. The total energy $E$, and the `anastrophy', defined as the spatial integral of $\Psi^2$, are known to be `rugged' invariants of this system \mycitep[e.g.,][]{shebalin2002statistical, biskamp_magnetohydrodynamic_2003}. Since $F_{2D,\Psi} = (F^+_{2D} - F^-_{2D})/2 \neq 0$, we expect energy to cascade to small scales, while anastrophy should undergo an inverse cascade \mycitep{fyfe_dissipative_1977}. This has not been observed in exploratory simulations with sub-box-scale forcing, possibly because $F_{2D}^\pm$ is non-trivially correlated with~$\bar{\zeta}_0^\pm$ via \cref{eq:barzeta1}. This constitutes an unusual situation where the forcing is `aware' of the phase of the field that it drives. A mismatch between the perpendicular scales of the forcing and the condensate may impact the conclusions of \S\cref{sec:discussion}, since the zonal shearing rate depends on the zonal wavenumber quadratically, as per \cref{eq:rates}. A full investigation is left for future work.}, and again assuming that $\zeta^+_n$ and $\zeta^-_n$ are of similar magnitude (as observed in the simulations), we find

\begin{equation} \label{eq:scalings}
    \bar{\zeta}_0 \sim \xi_0 \sim \bar{\zeta}_2 \sim \xi_2 \sim \frac{k_{\perp}^2 \bar{\zeta}_1^2}{\omega_\mathrm{Ad}} \sim \epsilon \bar{\zeta}_1 \sim \left(\frac{D^2}{k_{\perp}^2 \omega_\mathrm{Ad}} \right)^{1/3}.
\end{equation} From here on, $k_\perp$ should be understood to mean $k_{\perp \mathrm{A}}$.

It is perhaps surprising that the amplitudes of all of the undriven peaks scale identically. This might be expected in the condensate, which has no linear term, and hence constitutes a purely nonlinear subsystem. However, as $\epsilon \rightarrow 0$ the total field $\zeta_2$ does not tend towards the linear limit, but instead contains an additional peak at zero frequency. This could be interpreted as follows: as the turbulence becomes weaker and weaker, wavepackets become increasingly quasiparticle-like, so the conditions \cref{eq:qpmom} are more rigorously enforced. The wave modes, whilst themselves tending towards linearity, are ever more tightly entwined with the strongly nonlinear condensate. This leaves an indelible fingerprint of nonlinearity on the entire system, producing a state between the traditional paradigms of strong and weak turbulence. These non-perturbative scalings are not broken by retention of higher parallel wavenumbers, as discussed in the previous section. However, the time taken to reach the steady state does increase in systems with weaker nonlinearity, as discussed in \cref{app:transient}.

Crucially, the relation between the `zonal' amplitude and the pump waves is determined by \cref{eq:barzetazero}, in which the antenna does not appear. As such, the scaling from \cref{eq:scalings}

\begin{equation} \label{eq:scaling}
	\bar{\zeta}_0 \sim \frac{k_\perp^2 \bar{\zeta}_1^2}{\omega_\mathrm{Ad}},
\end{equation} is, in theory, independent of the nature of the pump-wave forcing. This means that in the saturated state, this relation might be expected to hold for more realistic TAE antennae, which would drive exponential growth of the mode until a threshold set by wave-particle trapping \mycitep{berk_saturation_1990}. As anticipated above, this assumes that the turbulent system does not acquire phases from the new antenna that destroy the scaling relations.

Finally, the balance of the Elsasser fields implies that `zonal' flows and currents (corresponding to the $\Phi$ and $\Psi$ fields) are generated equally in this model, as there is no additional physics included that could break this symmetry.

\subsection{Similar results in the literature} \label{sec:lit}

We note that \cref{eq:scaling} resembles Eq. (14) of \mycitet{barberis2025perturbative}, but is in fact a consequence of entirely different physics. Their result was obtained via energy conservation within a simple two-mode system where the zonal component has no dissipation, so is slaved to the pump wave at all times via a quadratic beat-driven mechanism. This mechanism is derived in \mycitet{chen_effects_2024} using non-MHD physics, and is assumed to be valid at all times, so that Eq. (14) of \mycitet{barberis2025perturbative} also holds at all times. In contrast, our scaling is certainly not valid during the transient phase of our system (see \cref{app:transient}). Finally, the constant of proportionality in \mycitet{barberis2025perturbative} is much smaller, containing an extra factor of $\omega_{\star i}/\omega_\mathrm{Ad}$, where $\omega_{\star i}$ is the ion diamagnetic frequency. This is a consequence of the beat-driven mechanism relying on a small imbalance between the Reynolds and Maxwell stresses.

Another estimate of the saturated zonal amplitude appears in \mycitet{yan_alfven_2025}. It is independent of TAE amplitude, because the TAEs are assumed to be strongly driven, so that they do not saturate via wave-particle trapping (as in the marginal case) but via flow shear from the zonals, which are themselves generated by the non-MHD mechanism of \mycitet{chen_effects_2024}. This produces a coupled set of equations, not unlike \crefrange{eq:barzetazero}{eq:barzeta2}, which have a fixed steady-state solution.

Finally, we refer back to \cref{eq:schekscale}, which first appeared in \mycitet{schekochihin_weak_2012}, and looks similar to \cref{eq:scalings}. As previously discussed, this utilised dimensional and stochastic arguments to justify an expression for the magnitude of the entire condensate (not separated by frequency) in terms of the $\zeta_1$ fields. There is an element of inevitability in this correspondence given the limited timescales and ordering parameters available in weak MHD turbulence, but the new result \cref{eq:scalings} is a more precise statement about the frequency structure of the condensate under coherent forcing.

\subsection{Generation and saturation of the `zonal' modes} \label{sec:mech}

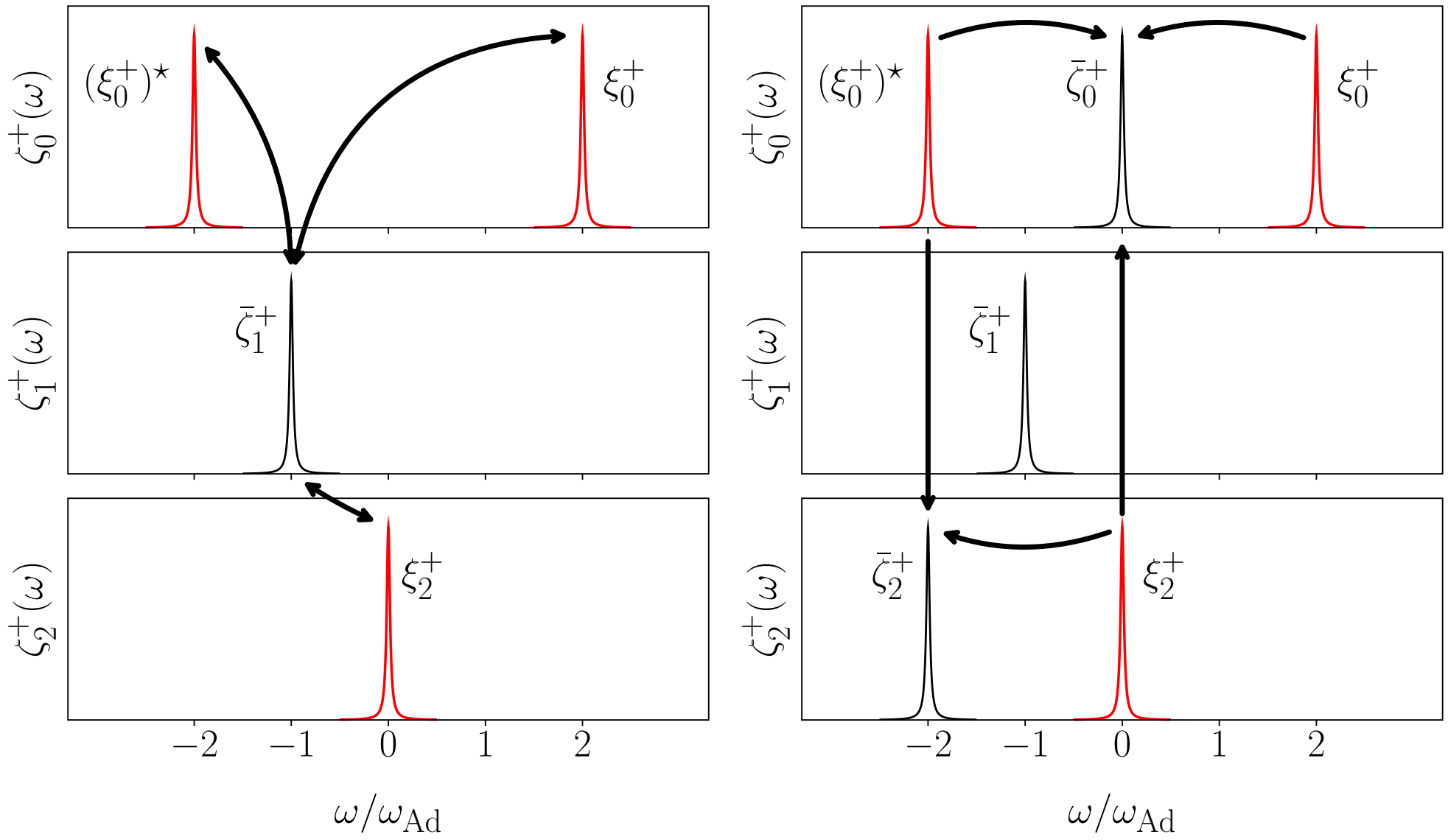
\begin{figure}
	\centering
	\pgfplotsset{
		flow schematic axis/.style={
			scale only axis,
			width=5.208551in,
			height=1.801959in,
			xmin=-3.3, xmax=3.3,
			ymin=0, ymax=1.15,
			xtick={-2,-1,0,1,2},
			ytick=\empty,
			axis lines=box,
			axis line style={line width=0.803pt},
			tick align=outside,
			tick style={black,line width=0.803pt},
			major tick length=0.048611in,
			xtick pos=bottom,
			tick label style={font=\sffamily\fontsize{24.88}{29.856}\selectfont,scale=0.924653},
			axis on top,
			clip=false,
		},
		flow schematic on shell/.style={samples=501, no marks, black, line width=1.2045pt},
		flow schematic off shell/.style={samples=501, no marks, red, line width=1.505625pt},
	}
	\tikzset{
		flow schematic text/.style={font=\sffamily\fontsize{24.88}{29.856}\selectfont,scale=0.924653},
		flow schematic arrow/.style={line cap=round,line join=round,line width=3.01125pt},
	}
	\resizebox{0.48\linewidth}{!}{%
		\begin{tikzpicture}[x=1in,y=1in]
			\path[use as bounding box] (0,0) rectangle (5.867346,6.962354);
			
			\begin{axis}[flow schematic axis, at={(0.558794in,5.060395in)}, anchor=south west, xticklabels=\empty]
				\addplot[flow schematic off shell, domain=-2.5:-1.5]
				{0.0004/((x+2)^2+0.0004)};
				\addplot[flow schematic off shell, domain=1.5:2.5]
				{0.0004/((x-2)^2+0.0004)};
			\end{axis}
			\begin{axis}[flow schematic axis, at={(0.558794in,3.059825in)}, anchor=south west, xticklabels=\empty]
				\addplot[flow schematic on shell, domain=-1.5:-0.5]
				{0.0004/((x+1)^2+0.0004)};
			\end{axis}
			\begin{axis}[flow schematic axis, at={(0.558794in,1.059255in)}, anchor=south west]
				\addplot[flow schematic off shell, domain=-0.5:0.5]
				{0.0004/(x^2+0.0004)};
			\end{axis}
			
			\node[flow schematic text,rotate=90,anchor=south] at (0.503239,5.961375) {$\zeta_0^+(\omega)$};
			\node[flow schematic text,rotate=90,anchor=south] at (0.503239,3.960805) {$\zeta_1^+(\omega)$};
			\node[flow schematic text,rotate=90,anchor=south] at (0.503239,1.960235) {$\zeta_2^+(\omega)$};
			\node[flow schematic text,anchor=north] at (3.163070,0.503239) {$\omega / \omega_\mathrm{Ad}$};
			\node[flow schematic text,anchor=north west] at (0.637712,6.470624) {$(\xi^+_0)^\star$};
			\node[flow schematic text,anchor=north west] at (4.867687,6.470624) {$\xi^+_0$};
			\node[flow schematic text,anchor=north west] at (1.868824,4.470054) {$\bar{\zeta}^+_1$};
			\node[flow schematic text,anchor=north west] at (3.226204,2.469484) {$\xi^+_2$};
			
			\draw[flow schematic arrow] (2.371214,4.784436) .. controls (2.352166,5.441723) and (2.125230,6.017004) .. (1.690405,6.510279);
			\draw[flow schematic arrow] (2.410449,4.868941) -- (2.371214,4.784436) -- (2.327151,4.866527);
			\draw[flow schematic arrow] (1.714254,6.420213) -- (1.690405,6.510279) -- (1.776767,6.475319);
			\draw[flow schematic arrow] (2.407172,4.780871) .. controls (2.670015,5.941598) and (3.395592,6.554715) .. (4.583903,6.620220);
			\draw[flow schematic arrow] (2.466214,4.852944) -- (2.407172,4.780871) -- (2.384938,4.871349);
			\draw[flow schematic arrow] (4.498403,6.657237) -- (4.583903,6.620220) -- (4.502990,6.574030);
			\draw[flow schematic arrow] (2.505230,2.972584) .. controls (2.669375,2.865573) and (2.840615,2.771475) .. (3.018952,2.690291);
			\draw[flow schematic arrow] (2.552283,2.892170) -- (2.505230,2.972584) -- (2.597794,2.961978);
			\draw[flow schematic arrow] (2.960370,2.762740) -- (3.018952,2.690291) -- (2.925844,2.686895);
		\end{tikzpicture}%
	}
	\resizebox{0.48\linewidth}{!}{%
		\begin{tikzpicture}[x=1in,y=1in]
			\path[use as bounding box] (0,0) rectangle (5.867346,6.962354);
			
			\begin{axis}[flow schematic axis, at={(0.558794in,5.060395in)}, anchor=south west, xticklabels=\empty]
				\addplot[flow schematic on shell, domain=-0.5:0.5]
				{0.0004/(x^2+0.0004)};
				\addplot[flow schematic off shell, domain=-2.5:-1.5]
				{0.0004/((x+2)^2+0.0004)};
				\addplot[flow schematic off shell, domain=1.5:2.5]
				{0.0004/((x-2)^2+0.0004)};
			\end{axis}
			\begin{axis}[flow schematic axis, at={(0.558794in,3.059825in)}, anchor=south west, xticklabels=\empty]
				\addplot[flow schematic on shell, domain=-1.5:-0.5]
				{0.0004/((x+1)^2+0.0004)};
			\end{axis}
			\begin{axis}[flow schematic axis, at={(0.558794in,1.059255in)}, anchor=south west]
				\addplot[flow schematic on shell, domain=-2.5:-1.5]
				{0.0004/((x+2)^2+0.0004)};
				\addplot[flow schematic off shell, domain=-0.5:0.5]
				{0.0004/(x^2+0.0004)};
			\end{axis}
			
			\node[flow schematic text,rotate=90,anchor=south] at (0.503239,5.961375) {$\zeta_0^+(\omega)$};
			\node[flow schematic text,rotate=90,anchor=south] at (0.503239,3.960805) {$\zeta_1^+(\omega)$};
			\node[flow schematic text,rotate=90,anchor=south] at (0.503239,1.960235) {$\zeta_2^+(\omega)$};
			\node[flow schematic text,anchor=north] at (3.163070,0.503239) {$\omega / \omega_\mathrm{Ad}$};
			\node[flow schematic text,anchor=north west] at (2.657998,6.470624) {$\bar{\zeta}^+_0$};
			\node[flow schematic text,anchor=north west] at (0.637712,6.470624) {$(\xi^+_0)^\star$};
			\node[flow schematic text,anchor=north west] at (4.867687,6.470624) {$\xi^+_0$};
			\node[flow schematic text,anchor=north west] at (1.868824,4.470054) {$\bar{\zeta}^+_1$};
			\node[flow schematic text,anchor=north west] at (1.079649,2.469484) {$\bar{\zeta}^+_2$};
			\node[flow schematic text,anchor=north west] at (3.289338,2.469484) {$\xi^+_2$};
			
			\draw[flow schematic arrow] (3.163070,2.737272) .. controls (3.163070,3.474603) and (3.163070,4.196406) .. (3.163070,4.902681);
			\draw[flow schematic arrow] (3.121403,4.819348) -- (3.163070,4.902681) -- (3.204737,4.819348);
			\draw[flow schematic arrow] (1.688787,6.606198) .. controls (2.145526,6.764802) and (2.587597,6.769896) .. (3.014998,6.621479);
			\draw[flow schematic arrow] (2.949944,6.688177) -- (3.014998,6.621479) -- (2.922607,6.609455);
			\draw[flow schematic arrow] (4.637353,6.606198) .. controls (4.180614,6.764802) and (3.738543,6.769896) .. (3.311142,6.621479);
			\draw[flow schematic arrow] (3.403533,6.609455) -- (3.311142,6.621479) -- (3.376196,6.688177);
			\draw[flow schematic arrow] (1.584721,4.949300) .. controls (1.584721,4.211969) and (1.584721,3.490166) .. (1.584721,2.783891);
			\draw[flow schematic arrow] (1.626388,2.867224) -- (1.584721,2.783891) -- (1.543054,2.867224);
			\draw[flow schematic arrow] (3.059004,2.587294) .. controls (2.602265,2.428690) and (2.160195,2.423596) .. (1.732794,2.572013);
			\draw[flow schematic arrow] (1.797847,2.505315) -- (1.732794,2.572013) -- (1.825184,2.584037);
		\end{tikzpicture}%
	}
	\caption{At very short times (left column), the linearly driven peak is excited directly. The off-shell peaks (in red) are slaved to these modes, and adjust their amplitudes instantaneously. At slightly later times (right column), the other on-shell peaks $\bar{\zeta}_0$ and $\bar{\zeta}_2$ (in black) emerge due to the drive by various combinations of the off-shell peaks. Note that the arrows are illustrative: in reality, interactions are between opposite Elsasser fields, and may also include negative values of~$k_\parallel$.}
	\label{fig:schematic_flow}
\end{figure}

As anticipated in \S\cref{sec:freqdoubl}, the characteristic frequencies of the condensate — described by \cref{eq:Schekcond} — appear to be $\pm 2 \omega_\mathrm{Ad}$. However, in our expansion, we allowed for the presence of an additional peak $\bar{\zeta}_0$ at $\omega=0$, which was shown to be of equal magnitude. Given that these are the `zonal' modes of interest, we now elucidate explicitly how they are generated. We can infer from \cref{eq:barzetazerofirst} or \cref{eq:barzetazero} that two mechanisms contribute equally, both at the second level of nonlinearity, as shown schematically in \cref{fig:schematic_flow}. By this we mean that the $\bar{\zeta}_0$ modes are nonlinearly beat-driven by the $\xi_0$ and $\xi_2$ modes, which are themselves beat-driven by the pump waves.

The first type of drive term present in \cref{eq:barzetazero} is due to $N \left[\xi_0^\mp, \xi_0^\pm\right]$. This is intuitive: returning to~\cref{eq:Schekcond}, we note that each of the Elsasser fields $\zeta_0^\pm$ is directly driven at the two frequencies $2 \omega_\mathrm{Ad}$ and $- 2 \omega_\mathrm{Ad}$, due to the $k_\parallel = k_{\parallel \mathrm{d}}$ and $k_\parallel = - k_{\parallel \mathrm{d}}$ contributions. The first term on the right-hand side of \cref{eq:Schekcond} is the self-advection of the condensate (but by opposite Elsasser fields), which allows the $2 \omega_\mathrm{Ad}$ and $- 2 \omega_\mathrm{Ad}$ components to interact, generating bands at multiples of $2 \omega_\mathrm{Ad}$, of which the largest amplitude is at $\omega = 0$. This is why the effective forcing at zero frequency is not quadratic in pump-wave amplitude but quartic, and occurs indirectly.

The second type of drive term in \cref{eq:barzetazero} is due to $N \left[\xi_2^\mp, (\xi_2^\pm)^\star \right]$. The off-shell component $\xi_2^\pm$ is at zero frequency, and is initially generated by the direct interaction of two pump waves with $k_\parallel= k_{\parallel \mathrm{d}}$, but of opposite Elsasser fields and hence opposite frequencies. These modes can then directly drive the condensate at $\omega=0$, naturally generating zero-frequency modes. Referring back to \cref{eq:Schekcond}, we see that the drive term hides a DC component due to the off-shell parts of the $\zeta^\pm_2$ fields.

Given that the generation of `zonal' modes is at the second level of nonlinearity, why then is this peak not small? In \cref{eq:barzetazero}, $\bar{\zeta}_0$ saturates by self-advection: if the field is small, the rate of energy transfer out of a given wavenumber is also small. This means that the amplitude builds until self-advection is sufficiently strong to balance the effective driving from $\bar{\zeta}_1$. An analysis of the transient behaviour in \cref{app:transient} shows that $\bar{\zeta}_0$ initially grows more slowly than $\bar{\zeta}_1$ and $\xi_0$, but saturates at the same time.

\section{Numerical verification} \label{sec:numver}

In \S\cref{sec:nummethods} we describe the code used in our simulations and the parameter scan used to verify the relations found in \S\cref{sec:analresults}. The results of this scan are then presented in \S\cref{sec:numresults}.

\subsection{Numerical methods}\label{sec:nummethods}

\begin{figure}
	\centering
	\includegraphics[width=0.6\linewidth]{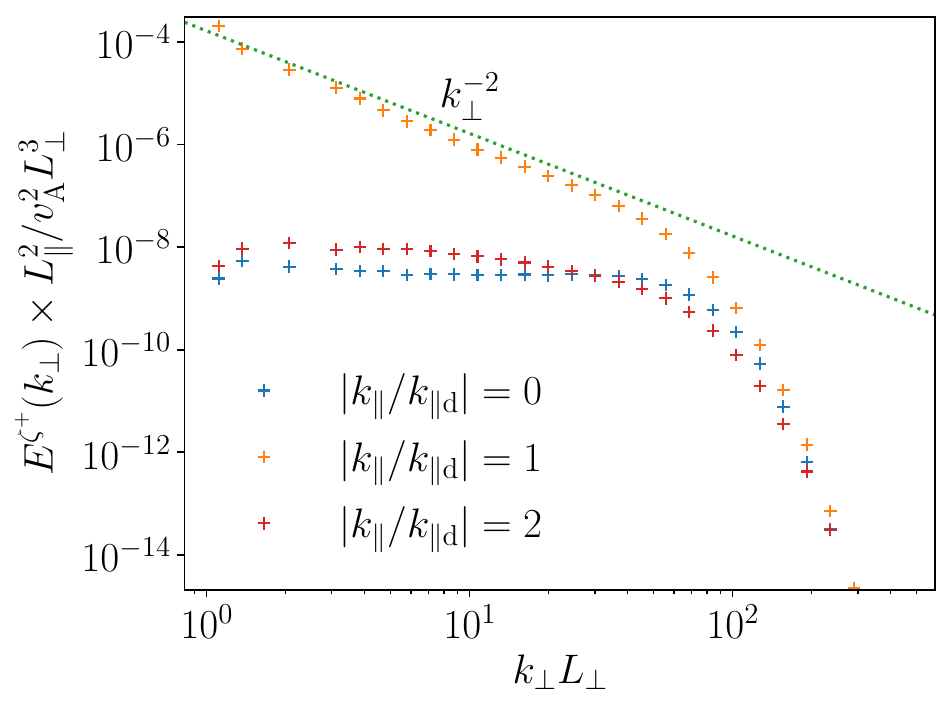}
	\caption{Perpendicular energy spectrum for $\Psi_\mathrm{A} = 2 \times 10^{-7} v_\mathrm{A} L_\perp^2 / L_\parallel$, showing the expected $k_\perp^{-2}$ scaling at low $k_\perp$ \mycitep[e.g.,][]{schekochihin_mhd_2022}. The spectrum then steepens on entering the dissipation range, which begins before the transition to strong turbulence. The latter transition is expected at $k_\perp L_\perp \sim 300$, where fit lines for different parallel wavenumbers — extrapolated from large scales — would meet in energy \mycitep{schekochihin_weak_2012}. We define $E^{\zeta^+} (k_\perp) = L_x L_y \sum_{\mathbf{k} \in B_\mathrm{bin}} k_\perp^3 |\zeta^+_\mathbf{k}|^2/2 \pi N_\mathrm{bin}$, where $B_\mathrm{bin}$ is a perpendicular wavenumber bin containing $N_\mathrm{bin}$ modes.}
	\label{fig:en_spectrum}
\end{figure}

For all simulations presented in this paper, dimensionless versions of \crefrange{eq:Psieq}{eq:lapphi} are solved pseudospectrally with the antenna of \S\cref{sec:antderiv} in a triply periodic box with $L_x = L_y = L_z$. The code used is a modified version of the codes employed in \mycitet{ivanov_zonally_2020} and \mycitet{zhang_runaway_2025}. The antenna is chosen to inject Alfvén waves at the box scale, and perpendicular fourth-order hyperdissipation is used to remove energy at small scales. No parallel dissipation is found to be necessary. Resolution is set to $(N_x, N_y, N_z) = (1024,1024,9)$, so that only $|k_\parallel|/k_{\parallel \mathrm{d}} = 0,1,2$ are retained after dealiasing, in accordance with \S\cref{sec:paralleltrunc}. To ensure the validity of this truncation, it is important to avoid entering the strongly turbulent regime, where parallel wavenumbers couple. This would always happen at small enough perpendicular scales \mycitep[e.g.,][]{schekochihin_mhd_2022}, so we adjust the hyperdissipation coefficient to ensure that the transition between the inertial range and the dissipation range occurs at lower $k_\perp$ than the transition to strong turbulence (an example perpendicular spectrum is shown in \cref{fig:en_spectrum}). Simulations are run for a period of $\sim 6 \tau_{\mathrm{nl}, 0}$, where $\tau_{\mathrm{nl},0} \equiv (k_\perp^2 \zeta_0)^{-1} \sim \epsilon^{-2} \tau_\mathrm{A}$ is the nonlinear time of the condensate. This ensures saturation of the system, as described in \cref{app:transient}.

It is found that the choice of the pump-wave wavevectors $\mathbf{k}_{\perp 1}$ and $\mathbf{k}_{\perp 2}$ is unimportant so long as the perpendicular polarisations of the two waves are not the same\footnote{Different transient behaviour is found if the magnitudes of the two perpendicular wavevectors are equal, $k_{\perp 1} = k_{\perp 2}$, as the nonlinearities in \cref{eq:lapphi} then initially cancel.}. To mimic TAE physics, $\mathbf{k}_{\mathrm{A}1}$ and $\mathbf{k}_{\mathrm{A}2}$ are set to be $(k_x L_x, k_y L_y, k_z L_z) = (1,0,1)$ and $(1,1,1)$, as the two components of a single TAE have the same toroidal and radial dependence, but are neighbouring poloidal harmonics. Seven simulations are carried out within a range of antenna amplitudes $\Psi_\mathrm{A} L_\parallel / v_\mathrm{A} L_\perp^2 = 2 \times 10^{-7} \text{--} 2 \times 10^{-4}$, spaced at equal logarithmic intervals. This ensures that $D$ (proportional to $\Psi_\mathrm{A}$) is varied across three orders of magnitude, so that the saturated pump-wave amplitudes and hence the weak-turbulence parameter $\epsilon$ span a decade ($\epsilon \simeq 0.007 \text{--} 0.07$). We define

\begin{equation} \label{eq:avzeta}
	\bar{\zeta}_{n, \mathrm{rms}}^+ \equiv \left[\sum_{\boldsymbol{k}_\perp}\left|\frac{1}{T_1}\int_{t_{\mathrm{eq}}}^{t_{\mathrm{eq}}+T_1}\zeta_{\boldsymbol{k}_\perp,nk_{\parallel\mathrm{d}}}^{+}(t')\,e^{-in\omega_{\mathrm{Ad}}t'}\,\mathrm{d}t'\right|^2\right]^{1/2}, \qquad n=0,1,
\end{equation} where $t_\mathrm{eq}$ is picked so that the system has reached steady state, and $T_1$ is as many multiples of $\tau_{\mathrm{nl},0}$ as memory allows ($ \sim 0.5 \text{--} 5$). Note that during this period, the fields must also be saved sufficiently often to resolve the oscillations at $2 \omega_\mathrm{Ad}$, otherwise the high-frequency structure would be aliased into the low-frequency peak. As $\tau_\mathrm{nl,0}$ is larger than $\tau_\mathrm{nl,1}$ by $1/\epsilon$, the frequencies that must be simultaneously resolved are separated by $\tau_\mathrm{A}/\tau_\mathrm{nl,0} \sim \epsilon^2$. In addition to the increasingly long time required to reach saturation for smaller $\epsilon$, this memory limit places a computational lower bound on the value of $\epsilon$ we can investigate. Ensuring that there is approximately a decade of inertial range sets the upper limit on $\epsilon$, as for stronger antennae, the dissipation range must begin at larger scales to avoid the transition to strong turbulence.

\subsection{Numerical results}\label{sec:numresults}

\begin{figure}
	\centering
	\includegraphics[width=0.6\linewidth]{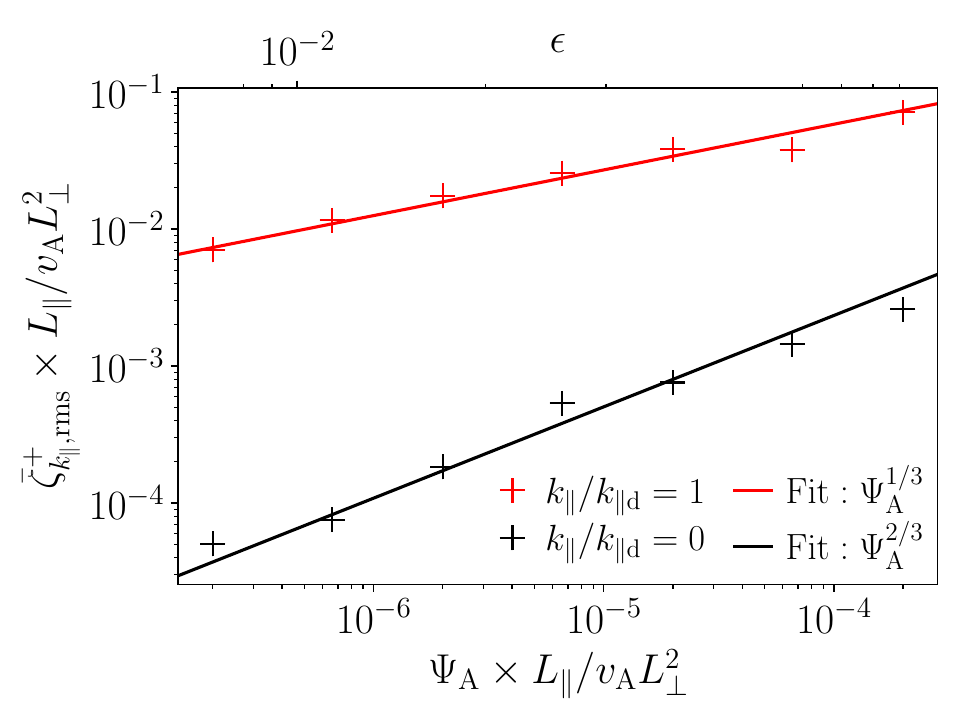}
	\caption{Variation of the fields $\bar{\zeta}^+_{0, \mathrm{rms}}$ and $\bar{\zeta}^+_{1, \mathrm{rms}}$, averaged using \cref{eq:avzeta}, with $\Psi_\mathrm{A}$ (proportional to $D$), along with the expected scaling laws given by \cref{eq:scalings}. The weak-turbulence parameter $\epsilon$ defined in \cref{eq:epsilondefn} (which is proportional to $\bar{\zeta}_1^+$) is derived from the data and plotted on the top axis.}
	\label{fig:scan_D}
\end{figure}

\begin{figure}
	\centering
	\includegraphics[width=0.6\linewidth]{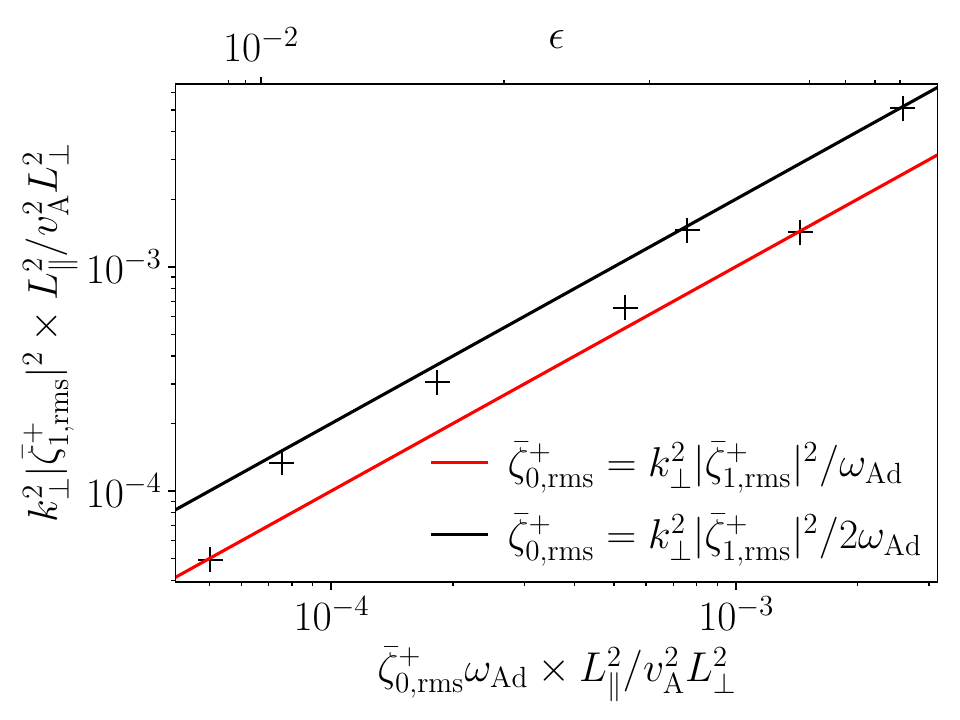}
	\caption{Amplitude of the `zonal' modes as a function of the pump-wave amplitude, averaged according to \cref{eq:avzeta}. The scaling  \cref{eq:scaling} is confirmed, with the order-unity constant found to be between $0.5$ and $1$. The weak-turbulence parameter $\epsilon$ is again plotted on the top axis, as defined in \cref{eq:epsilondefn}.}
	\label{fig:scan_ratio}
\end{figure}

We now compare the saturated amplitudes measured in the simulations with the analytical results derived in \S\cref{sec:analresults}. Good agreement is found with the predicted scalings~\cref{eq:scalings} of $\bar{\zeta}_0^\pm$ and $\bar{\zeta}_1^\pm$ with antenna amplitude, as shown in \cref{fig:scan_D}. Of the scalings derived, the relationship between the amplitudes of the pump waves and the `zonal' modes — given by \cref{eq:scaling} — is of greatest interest. This is verified directly in \cref{fig:scan_ratio}.

Error bars are omitted in \cref{fig:scan_D,fig:scan_ratio} for a number of reasons. The most obvious error estimate would be the variance of the field over the averaging period. This is minuscule for the weakest-driven cases, where the nonlinear time is vastly greater than the Alfvén time ($\tau_{\mathrm{nl,0}} \sim 10^5 \tau_\mathrm{A}$). For these cases, the more significant error is set by the finite averaging period $T_1$, which is constrained by the available computational resources to be $T_1 \lesssim \tau_{\mathrm{nl},0}$. This means that the time average does not fully eliminate the order-unity oscillations of the field over the timescale~$\tau_{\mathrm{nl},0}$, which therefore dominate the error in $\bar{\zeta}_0$. In any case, the equations \cref{eq:scalings} are just scalings, not exact expressions, and all data points are well within an order of magnitude of the fit. The close correspondence between the analytical and numerical results is presented as evidence of the validity of our reduced system of equations, and hence of the parallel-wavenumber truncation and frequency separation carried out in \S\cref{sec:analmethods}. Good agreement is also found for the transient behaviour of $\bar{\zeta}_0$, $\tilde{\zeta}_0$ and $\bar{\zeta}_1$, which is shown in \cref{app:transient}. As discussed in \S\cref{sec:analresults}, it appears that the system does not develop correlations in the phases that destroy the scalings \cref{eq:scalings} — at least for this choice of the driving antenna.

\section{Discussion} \label{sec:discussion}

The results presented above have been derived within a model lacking equilibrium gradients, complex geometry and finite-Larmor-radius effects, which are all present in a realistic fusion device. In a more comprehensive plasma model, the terms responsible for zonal-flow generation in the current system should be unaffected, especially since continuum damping should not influence the important modes. These live in the TAE gap at $\omega_\mathrm{Ad}$, in the ellipticity-induced Alfvén eigenmode (EAE) gap \mycitep{betti_ellipticity_1991} at $2\omega_\mathrm{Ad}$, or at zero frequency, where they are expected to interact with low-frequency microturbulence.  However, the inclusion of the aforementioned additional physics will generate new terms. For example, \mycitet{chen_effects_2024} show that zonal modes can be driven quadratically (and hence `directly') in the pump-wave amplitude, albeit with a much smaller prefactor than the quartic drive presented here. This is possible when compressibility, geometry, or finite-Larmor-radius effects introduce a small imbalance between the Maxwell and Reynolds stresses of the shear Alfvén wave. Given that pump-wave amplitudes for TAEs are small, whether a quadratic term with a small prefactor would dominate over a quartic term will require careful investigation, as will the nature of more realistic damping terms for the zonal modes.

 Presuming for now that we can extrapolate our results to real tokamaks, let us ask whether the zonal flows produced in the way that we have proposed would be sufficiently strong to alter transport in the core. The key parameter is the ratio of the zonal $\mathbf{E} \times \mathbf{B}$ shearing rate to the radial decorrelation rate of the curvature-driven ITG turbulence, which is often estimated by the ITG growth rate \mycitep{waltz_toroidal_1994}. When this ratio approaches unity, transport can be strongly suppressed in a single flux tube. The two rates can be approximated by

\begin{equation} \label{eq:rates}
    \omega_\mathrm{E} = \frac{\mathrm{d}}{\mathrm{d}r} v_{\mathrm{E}, y} \sim k_r^2 \bar{\Phi}_\mathrm{ZF}, \, \, \,
    \gamma_\mathrm{ITG} \sim k_y \rho_i \frac{c_s}{\sqrt{L_T L_B}},
\end{equation} where $k_r$ is the approximate radial wavenumber of the zonal flow, $k_y$ is the wavenumber of the ITG turbulence in the binormal direction, $\rho_i$ is the ion Larmor radius, and $L_T$ and $L_B$ are the temperature-gradient scale and the magnetic-field scale, respectively. Using $|\delta u_\perp|/v_\mathrm{A} \sim |\delta B_\perp|/B$, our saturation estimate \cref{eq:scalings} for $\bar{\Phi}_\mathrm{ZF}$, produced by a TAE of frequency $\omega_\mathrm{TAE}$ and amplitude $\Phi_\mathrm{TAE}$, can be expressed as

\begin{equation}
    \bar{\Phi}_\mathrm{{ZF}} \sim \frac{k_\perp^2}{\omega_\mathrm{TAE}} \bar{\Phi}_{\mathrm{TAE}}^2 \sim  \left[\frac{v_\mathrm{A}}{k_{\parallel}} \left( \frac{\delta B_\perp}{B} \right)^2 \right]_\mathrm{TAE},
\end{equation} which gives (assuming $k_r$ is the same for the zonal mode and the TAE)

\begin{equation} \label{eq:ratio}
    \frac{\omega_\mathrm{E}}{\gamma_\mathrm{ITG}} \sim \beta^{-1/2} \left[ k_{y} \rho_i \right]^{-1}_\mathrm{ITG} \left[ k_r \sqrt{L_T L_B} \, \frac{k_r}{k_{\parallel}}  \left( \frac{\delta B_\perp}{B} \right)^2 \right]_\mathrm{TAE}.
\end{equation}

As $\delta B_\perp / B$ is very small for TAEs, it is evident that the other factors in this expression must accumulate to a large number to appreciably modify transport. We stress, however, that a significant effect cannot be ruled out a priori for future reactors, which will have high-$n$ `local' TAEs \mycitep{cheng_high-n_1985}. As these are formally gyrokinetic fluctuations, the right-hand side of \cref{eq:ratio} is $\mathcal{O}(1)$ in $\rho_{\star i} = \rho_i/a$ (where $a$ is the minor radius of the device), so the zonal shear is not \textit{a priori} negligible. Proceeding more generally, for ITG turbulence, we estimate $k_y \rho_i \sim L_T/qR$ \mycitep{barnes_critically_2011}. Furthermore, $k_\parallel$ for the TAE can be estimated as $1/qR$, while the radial extent of the driven mode must approximately match the distance that alpha particles drift away from the flux surface over a poloidal transit time, to allow for efficient energy exchange. This width (for passing alphas) is given by $q$ multiplied by their Larmor radius at the Alfvén resonance, namely $q v_A/\Omega_\alpha$, where $\Omega_\alpha$ is the cyclotron frequency of the alpha particles. Therefore, for a TAE of given amplitude, \cref{eq:ratio} becomes

\begin{equation} \label{eq:shearing_ratio}
    \frac{\omega_{E}}{\gamma_\mathrm{ITG}} \sim \beta^{1/2}  \left( \frac{R}{a} \right)^{5/2} \sqrt{\frac{a}{L_T}} \left( \frac{Z_\alpha m_i}{Z_i m_\alpha} \right)^2 \rho_{\star i}^{-2} \left( \frac{\delta B_\perp}{B} \right)^2_\mathrm{TAE},
\end{equation} where we have assumed $L_B \sim R$, and $Z_s$ denotes the charge of species $s$. To simplify this expression further, we approximate $L_T \sim a$, $Z_\alpha m_i / Z_i m_\alpha \approx 1$, and note that, for a wide range of devices, $\beta^{1/2} (R/a)^{5/2}$ can be found numerically to be order unity. We can therefore reduce the condition $\omega_{E}/\gamma_\mathrm{ITG} \gtrsim 1$ to

\begin{equation} \label{eq:simplified}
    \left( \frac{\delta B_\perp}{B} \right)_\mathrm{TAE} \gtrsim  \rho_{\star i},
\end{equation} the size at which TAE-generated magnetic perturbations can have an appreciable effect on transport. We conclude, therefore, that in order to affect ion-scale gyrokinetic turbulence, TAEs simply require gyrokinetic-sized amplitudes.

We now briefly comment on whether such amplitudes are likely to cause unacceptable alpha-particle losses. This — bearing in mind the simplifications present in the model — aims to answer analytically whether there is a range of amplitudes within which the presence of TAEs can ever be beneficial in a tokamak. For current devices, with a small number of low-$n$ TAEs, we can estimate the threshold for unacceptable alpha-particle transport using the overlap criterion of the wave-particle resonances \mycitep{berk_map_1993}. The latest criterion derived by this method \mycitep{blyth_phd_inprep} can be approximated by

\begin{equation}
	\frac{\delta B}{B} \gtrsim \frac{Z_i m_\alpha}{Z_\alpha m_i} \rho_{\star i} \frac{a}{R} \beta^{-1/2} \frac{1}{4qs} \sim \rho_{\star i},
\end{equation} where we have assumed that the safety factor $q$ and the magnetic shear $s$ are order unity. This would imply that if a range of `beneficial' TAE amplitudes exists, it may be narrow. For future devices, where many TAEs will be present, resonance overlap may occur before the onset of large alpha-particle transport. Numerical work suggests that TAE amplitudes as large as $\delta B/B = 1 \times 10^{-2}$ may be permissible on ITER \mycitep{fitzgerald_predictive_2016}, significantly greater than the threshold for improved thermal confinement, given that $\rho_{\star i} \approx 2 \times 10^{-3}$ for that device. This suggests that there may be an opportunity to improve the core transport of fusion reactors by increasing TAE amplitudes.
 
\section{Conclusion} \label{sec:conclusion}

In this paper, we have assessed the strength of low-frequency flows generated by shear Alfvén pump waves in the simplest possible paradigm of reduced MHD in a uniform slab. Using a reduced system of equations, we have found that modes with $k_\parallel = 0$ and $\omega = 0$ are generated at the second level of nonlinearity, with transient and saturated behaviour verified numerically. This is a new mechanism, as it was previously believed that stationary flows could not be generated by shear Alfvén waves in such a simple model. 

Although our results may also be of fundamental interest for the theory of weak MHD turbulence, the key `practical' feature of this study is a prediction of the saturated amplitude of `zonal' modes for a given pump-wave amplitude. This allows an \textit{a priori} analytical estimate of whether TAEs can benefit magnetic fusion devices by suppressing thermal transport due to ITG turbulence in the core, without causing unacceptable alpha-particle losses. When applied to tokamaks, such an estimate suggests that current devices may have a narrow or non-existent range of `beneficial' TAE amplitudes, but that reactor-class devices such as ITER might have improved fusion performance if TAE amplitudes are increased. Future work is required to assess what may change with the inclusion of additional physics present in realistic machines.

\section*{Acknowledgements}

The authors would like to thank W. Clarke, R. Dutta, P. Ivanov, A. Wierzchucka, Y. Zhang, and the rest of the Oxford Plasma Theory group for useful discussions. Claude Opus 5.5 assisted the authors in proofreading the manuscript. The work of J. M. E. was funded by an Engineering and Physical Sciences Research Council (EPSRC) CASE studentship in partnership with the UK Atomic Energy Authority. D. A. U., J. M. E., and M. F. were supported by the EPSRC Fusion Grant 2022/27 (grant number EP/W006839/1), while J. R. R. was funded by grant number EP/W026341/1. To obtain further information on the data and models underlying this paper, please contact PublicationsManager@ukaea.uk. The work of A. A. S. was supported in part by the Simons Foundation via a Simons Investigator Award.

\section*{Declaration of interests}

The authors report no conflict of interest.

\appendix

\section{Short-time behaviour of the fields} \label[appendix]{app:transient}

The model discussed in this paper is not suitable for a description of TAE-driven fields in magnetic fusion devices at short times. Notwithstanding the lack of equilibrium gradients and geometry, the antenna does not inject energy exponentially like a realistic fast-particle drive, and the pump waves saturate when the `wave-wave' nonlinearity balances a constant driving term. This is very different from the paradigm established by \mycitet{berk_saturation_1990} for marginally unstable TAEs. For completeness, here we check that our numerical results match the predictions of the reduced model at short times.

\subsection{Short-time behaviour} \label{sec:shorttime}

\begin{figure}
    \centering
    \includegraphics[width=0.6\linewidth]{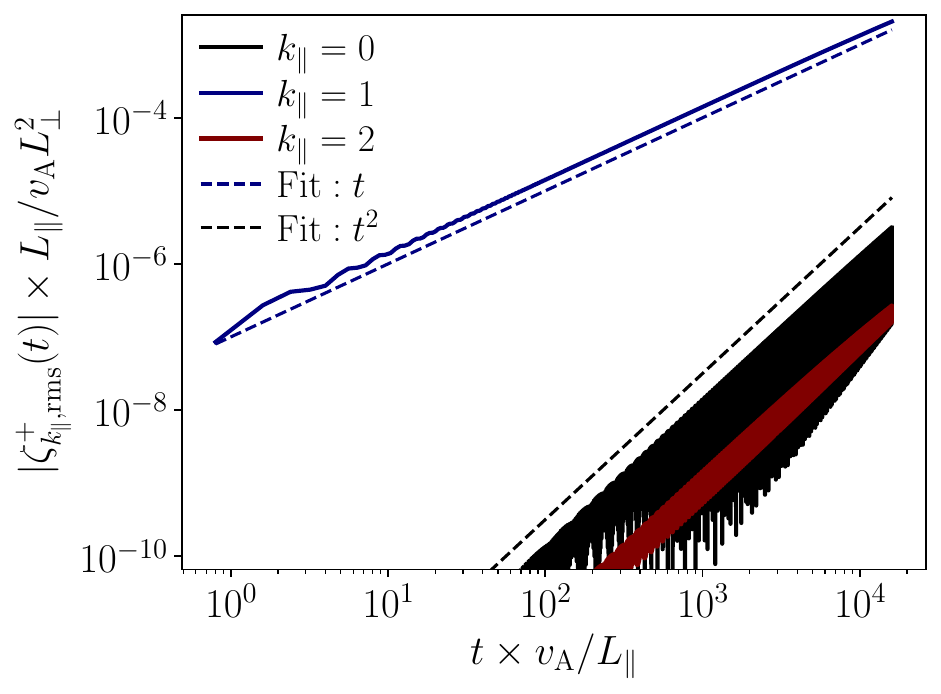}
    \caption{Growth of the total fields at short times for $\Psi_\mathrm{A} = 2 \times 10^{-7} v_\mathrm{A} L_\perp^2 / L_\parallel$. Good agreement is found with the expected $t$ and $t^2$ scalings predicted by \cref{eq:shorttimes_first} and \cref{eq:shorttimes_middle} respectively, noting that the $\bar{\zeta}_0$ and  $\bar{\zeta}_2$ contributions, proportional to $t^{5}$ according to \cref{eq:shorttimes_last}, are subdominant in this early phase. Note also that the $\zeta_2$ field is hardly oscillating because, unlike $\zeta_0$, the dominant component at short times — $\xi_2$ — is at zero frequency.}
    \label{fig:growthrates}
\end{figure}

\begin{figure}
    \centering
    \includegraphics[width=0.6\linewidth]{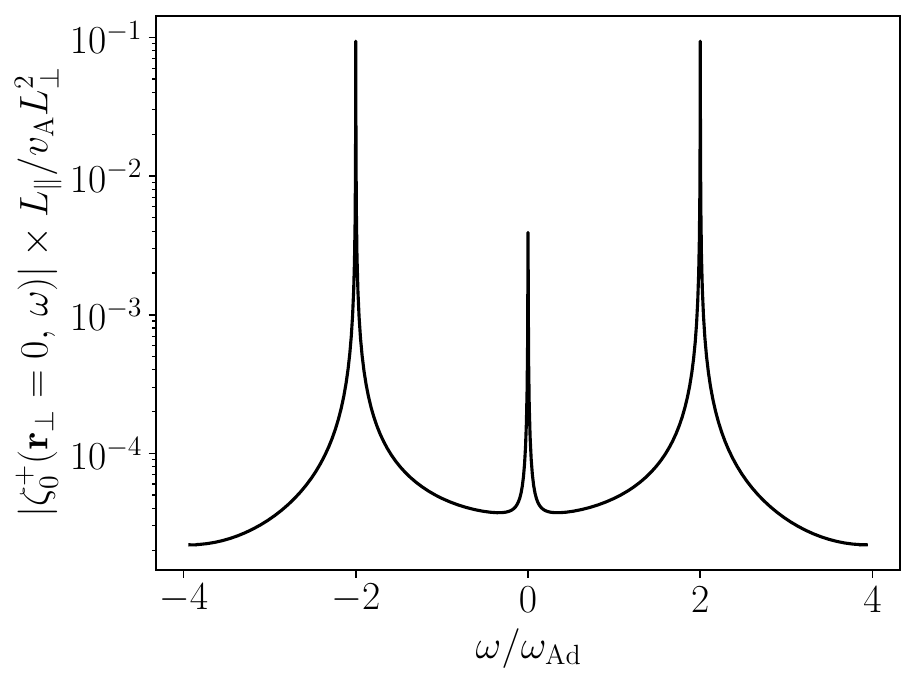}
    \caption{At short times in the condensate, the $\bar{\zeta}_0$ peak grows as $\sim D^4 t^5$ [see \cref{eq:shorttimes_last}], whereas the off-shell peaks grow as $\sim D^2 t^2$ [see \cref{eq:shorttimes_middle}]. Hence for times before $t_\mathrm{sat}$, the central peak is lower. This figure was produced using $\Psi_\mathrm{A} = 2 \times 10^{-6} v_\mathrm{A} L_\perp^2 / L_\parallel$, using the time series from $t v_\mathrm{A}/L_\parallel = 0 \text{--} 8000 $, where $t_\mathrm{sat} v_\mathrm{A}/L_\parallel \sim 10^4 $.}
    \label{fig:smallcenter}
\end{figure}

An analysis of equations \crefrange{eq:barzetazero}{eq:barzeta2} at short times yields the following relations:

\begin{align}
	& \bar{\zeta}_1
	\sim Dt, \label{eq:shorttimes_first} \\
	& \xi_0 \sim \xi_2
	\sim \frac{k_\perp^2}{\omega_\mathrm{Ad}} D^2 t^2, \label{eq:shorttimes_middle} \\
	& \bar{\zeta}_0 \sim \bar{\zeta}_2
	\sim \frac{k_\perp^6 D^4}{\omega_\mathrm{Ad}^2} t^5  \label{eq:shorttimes_last}.
\end{align} The off-shell, high-frequency parts of the condensate ($\xi_0$) grow proportionally to the pump wave squared, as shown in \cref{fig:growthrates}. This is a consequence of \cref{eq:xi0eqn}, and aligns with the `beat-driven' mechanism suggested by \mycitet{todo_nonlinear_2010} [whose Eq. (26) can be considered a version of \cref{eq:Schekcond}]. In our model, these off-shell peaks are as important as the `zonal' modes in saturating the pump waves, and dominate the condensate at short times. However, the zero-frequency modes ($\bar{\zeta}_0$) that can interact with ITG turbulence initially grow much more slowly. An example of a time at which the low-frequency part of the condensate has not yet caught up with the high-frequency part is shown in \cref{fig:smallcenter}. Note that for an antenna driving exponential growth of the pump wave at $\gamma_{\mathrm{TAE}}$, the condensate overall would grow at $2\gamma_{\mathrm{TAE}}$ while the `zonals' would grow at~$4 \gamma_{\mathrm{TAE}}$ from a much smaller initial amplitude.

\subsection{Saturation times}\label{sec:sattimes}

Using the previously derived expressions \cref{eq:scalings} for the saturated amplitudes, and matching them to \crefrange{eq:shorttimes_first}{eq:shorttimes_last}, we find that the saturation times have the same scaling for all fields:

\begin{equation}
	t_{0,\mathrm{sat}} \sim t_{1,\mathrm{sat}} \sim t_{2,\mathrm{sat}} \sim \left( \frac{\omega_\mathrm{Ad}}{D^2 k_\perp^4} \right)^{1/3}.
\end{equation}This is the same as the nonlinear time

\begin{equation}
    \tau_{\mathrm{nl},0} \sim \left( k_\perp^2 \bar{\zeta}_{0,\mathrm{sat}} \right)^{-1} \sim \left( \frac{\omega_\mathrm{Ad}}{D^2 k_\perp^4} \right)^{1/3}.
\end{equation} In terms of tokamak-relevant parameters, it can be estimated as follows, in the same vein as the calculation in \S\cref{sec:discussion}:

\begin{equation}
\begin{split}
    t_\mathrm{sat} \sim \frac{\omega_\mathrm{Ad}}{k_\perp^4 \zeta_{1, \mathrm{sat}}^2} &\sim \left[ \left( \frac{k_\perp^4}{k_r^2} \right)\frac{v_\mathrm{A}}{k_{\parallel}} \left( \frac{\delta B_\perp}{B} \right)^2 \right]^{-1}_\mathrm{TAE} \\
    &\sim \beta^{-1} \left(  \frac{R}{a} \right)^{-2} \rho_{\star i}^{2} \left(  \frac{Z_i m_\alpha}{Z_\alpha m_i} \right)^{2} \omega_\mathrm{TAE}^{-1}   \left(  \frac{\delta B_\perp}{B} \right)_\mathrm{TAE}^{-2}.
\end{split}
\end{equation} On JET, where $\rho_{\star i} \approx 4 \times 10^{-3}$, and a typical TAE has a frequency close to $200 \, \mathrm{kHz}$ and an amplitude of $\delta B_\perp / B \sim 10^{-3}$, the estimated saturation time is of order $0.1 \, \mathrm{ms}$, which is much shorter than typical pulse lengths. However, we again stress that the expression derived here will be different to that of a more realistic antenna, which would drive exponential growth in the TAE-like modes.

\bibliographystyle{jpp}

\bibliography{paper0}

\end{document}